%% file: main.tex
\documentclass[conference]{IEEEtran}
\usepackage{balance}
\usepackage{booktabs}
\usepackage{multirow}
\usepackage{booktabs}
\usepackage{colortbl}
\usepackage{array}
\usepackage{graphicx}
\usepackage{xcolor}
\usepackage[normalem]{ulem} 
\let\labelindent\relax
\usepackage{enumitem}
\usepackage{enumitem}
\usepackage[utf8]{inputenc} %
\usepackage[T1]{fontenc}    %
\usepackage{hyperref}       %
\usepackage{url}            %
\usepackage{amsfonts}       %
\usepackage{nicefrac}       %
\usepackage{microtype}      %
\usepackage{xcolor}         %
\usepackage{xspace}
\usepackage[normalem]{ulem}
\usepackage{wrapfig}
\usepackage{blindtext}
\usepackage[multiple]{footmisc}
\usepackage{dsfont}
\usepackage{caption}
\usepackage{bm}
\usepackage{mathrsfs}
\usepackage{xfrac}
\usepackage{algorithm}
\usepackage{algpseudocode}
\usepackage{ulem}
\usepackage{color,soul}
\usepackage{threeparttable}
\usepackage{makecell}

\newtheorem{corollary}{Corollary}

\newcommand{\uit}[1]{\uline{\textit{#1}}}
\newcommand{\cifar}{\textsc{Cifar-10}\xspace}
\newcommand{\gtsrb}{\textsc{Gtsrb}\xspace}
\newcommand{\mnist}{\textsc{Mnist}\xspace}
\newcommand{\contranet}{ContraNet\xspace}

\newcommand{\imagenet}{\textsc{ImageNet}\xspace}
\newcommand{\cinic}{\textsc{Cinic}\xspace}

\newcommand{\imagenetten}{\textsc{ImageNet10}\xspace}

\newcommand{\ssim}{\mathcal{SSIM}\xspace}
\newcommand{\G}{$\mathbf{G}$\xspace}
\newcommand{\E}{$\mathbf{E}$\xspace}
\newcommand{\dmm}{$\mathbf{DMM}$\xspace}
\newcommand{\dis}{\bm{$Dis$}\xspace}
\newcommand{\btsc}{\textsc{Btsc}\xspace}
\makeatletter
\def\hlinew#1{%
  \noalign{\ifnum0=`}\fi\hrule \@height #1 \futurelet
\reserved@a\@xhline}
\makeatother

\usepackage{cite}
\ifCLASSINFOpdf
  \graphicspath{{./figures/1.introduction/}{../jpeg/}}
  \DeclareGraphicsExtensions{.pdf,.jpeg,.png}
\else
  \DeclareGraphicsExtensions{.pdf,.jpeg,.png}
\fi
\usepackage[cmex10]{amsmath}
\usepackage{array}
\usepackage{eqparbox}
\usepackage[tight,footnotesize]{subfigure}
\usepackage{hyperref}
\usepackage{color,xcolor}
\usepackage{pifont}
\definecolor{mygreen}{RGB}{0 139 69}
\definecolor{mygreen2}{RGB}{0 205 0}
\definecolor{myred}{RGB}{205 38 38}
\definecolor{TartOrange}{HTML}{ff2e35}
\definecolor{Orange}{HTML}{ff7825}
\definecolor{Mango}{HTML}{ffc013}
\definecolor{AppleGreen}{HTML}{7cb81b}
\definecolor{Blue}{HTML}{1173b0}
\definecolor{BdazzledBlue}{HTML}{2e58a5}
\definecolor{Purple}{HTML}{5b3590}
\definecolor{Sunglow}{HTML}{FFCA3A}
\hypersetup{
	colorlinks=true,
	urlcolor=violet,
	citecolor=mygreen2,
}

\ifCLASSINFOpdf
\else
\fi
\begin{document}
%
\title{What You See is Not What the Network Infers: \\
Detecting Adversarial Examples Based on Semantic Contradiction 
\vspace{-1cm}
}


\author{
    \IEEEauthorblockN{
        Yijun Yang\IEEEauthorrefmark{1},
        Ruiyuan Gao\IEEEauthorrefmark{1},
        Yu Li\IEEEauthorrefmark{1},
        Qiuxia Lai\IEEEauthorrefmark{2} and
        Qiang Xu\IEEEauthorrefmark{1}
    }
    \IEEEauthorblockA{
       \IEEEauthorrefmark{1}\underline{CU}hk \underline{RE}liable Computing Laboratory (CURE Lab.),
       Dept. of Computer Science and Engineering,\\
       \textit{The Chinese University of Hong Kong}, Hong Kong S.A.R., China\\
       \{yjyang, rygao, yuli, qxu\}@cse.cuhk.edu.hk\\
        \IEEEauthorrefmark{2}State Key Laboratory of Media Convergence and Communication, \textit{Communication University of China}, Beijing, China\\
    }
}

\IEEEoverridecommandlockouts
\makeatletter\def\@IEEEpubidpullup{6.5\baselineskip}\makeatother
\IEEEpubid{\parbox{\columnwidth}{
    Network and Distributed Systems Security (NDSS) Symposium 2022\\
    27 February - 3 March 2022, San Diego, CA, USA\\
    ISBN 1-891562-74-6\\
    https://dx.doi.org/10.14722/ndss.2022.24001\\
    www.ndss-symposium.org
}
\hspace{\columnsep}\makebox[\columnwidth]{}}

\maketitle

\begin{abstract}

Adversarial examples (AEs) pose severe threats to the applications of deep neural networks (DNNs) to safety-critical domains, e.g., autonomous driving. While there has been a vast body of AE defense solutions, to the best of our knowledge, they all suffer from some weaknesses, e.g., defending against only a subset of AEs or causing a relatively high accuracy loss for legitimate inputs. Moreover, most existing solutions cannot defend against adaptive attacks, wherein attackers are knowledgeable about the defense mechanisms and craft AEs accordingly.

In this paper, we propose a novel AE detection framework based on the very nature of AEs, i.e., their semantic information is inconsistent with the discriminative features extracted by the target DNN model. To be specific, the proposed solution, namely \emph{\contranet}\footnote{Our code and models are available at \texttt{\url{https://github.com/cure-lab/ContraNet.git}}.}, models such contradiction by first taking both the input and the inference result to a generator to obtain a synthetic output and then comparing it against the original input. For legitimate inputs that are correctly inferred, the synthetic output tries to reconstruct the input. On the contrary, for AEs, instead of reconstructing the input, the synthetic output would be created to conform to the wrong label whenever possible. Consequently, by measuring the distance between the input and the synthetic output with metric learning, we can differentiate AEs from legitimate inputs. We perform comprehensive evaluations under various AE attack scenarios, and experimental results show that \contranet outperforms existing solutions by a large margin, especially 
under adaptive attacks. Moreover, our analysis shows that successful AEs that can bypass \contranet tend to have much-weakened adversarial semantics. We have also shown that \contranet can be easily combined with adversarial training techniques to achieve further improved AE defense capabilities.

\end{abstract}
\input{1.introduction.tex}

\input{2.background_relatedworks.tex}

\input{3.ContraNet_framework.tex}

\input{4.one_implementation.tex}

\input{6.whitebox_attack_evaluation.tex}
\input{7.adaptive_attack_evaluation.tex}

\input{9.limitation_and_futurework.tex}
\input{8.conclusion_and_futureworks.tex}

\section*{Acknowledgements}

We would like to acknowledge the contributions of Miss Bo Luo for her early exploration of this topic, and we thank the anonymous reviewers for their valuable comments.

This work was supported in part by General Research Fund of Hong Kong Research Grants Council (RGC) under Grant No. 14203521, No. 14205420, and No. 14205018.

\IEEEtriggercmd{\enlargethispage{-5in}}
\bibliographystyle{IEEEtranS}
\bibliography{ref.bib}

\input{appendix.tex}

\end{document}

%% file: 1.introduction.tex
\section{Introduction}
\label{sec:introduction}
Deep learning-based systems have achieved unprecedented success in numerous long-standing machine learning tasks~\cite{he2016deep, krizhevsky2012imagenet}. Their safety and trustworthiness have become public concerns with the broader deployment of Deep Neural Networks (DNNs) in various mission-critical applications, e.g., autonomous driving~\cite{automobile} and medical diagnosis~\cite{naren2021iomt,kaissis2020secure}. 
In these applications, incorrect decisions or predictions could cause catastrophic financial damages or even life losses~\cite{hirano2021universal}.

One of the primary threats to DNNs is Adversarial Examples (AEs),
which introduce subtle malicious perturbation to the inputs to fool the DNN model~\cite{biggio2013evasion,szegedy2013intriguing}.
While the perturbation is small and often imperceptible to humans, it dramatically changes the features extracted by the targeted DNN model, leading to wrong inference results. 
Depending on attackers' knowledge, adversarial attacks can be classified into three categories: 
\textit{black-box attack}, \textit{white-box attack}, and \textit{adaptive attack}. Their detailed descriptions are provided in Table~\ref{tab:typical_attack_scenario}. 

\begin{table*}[t]
	\centering
	\caption{Adversarial Attack Types}\label{tab:typical_attack_scenario}
		\setlength{\tabcolsep}{5mm}{
	\begin{tabular}{cc}

		\hlinew{1pt}
		\multicolumn{1}{c}{\textbf{Attack Type}} & \textbf{Description}            \\ \hline
		\rowcolor[HTML]{EFEFEF} 
		\textbf{Black-box attack}                            & \textit{Adversaries have no access to either the detailed information of the target model or its defense mechanism, but can query the model}~\cite{papernot2017practical}.                    \\
		\rowcolor[HTML]{C0C0C0} 
		\textbf{White-box attack}                            & \textit{The model architecture and parameters are exposed to the adversaries, but the defense mechanism is kept confidential (non-adaptive)}~\cite{ling2019deepsec, CarliniEvaluatingAdversarialRobustness2019}.                           \\
		\rowcolor[HTML]{EFEFEF} 
		\textbf{Adaptive attack}                             & \textit{Adversaries have full knowledge of both the target model and the defense mechanism, and could craft attacks accordingly}~\cite{CarliniEvaluatingAdversarialRobustness2019}. \\
		\hlinew{1pt}
		\end{tabular}}
	\end{table*}

A vast body of research has been dedicated to AE defense, considering the severity of the threat. Existing methods include model robustification with adversarial training techniques (e.g.,~\cite{madry-PGD,shafahi2019adversarial}), input transformation to mitigate the impact of AEs (e.g.,~\cite{MengMagNetTwoProngedDefense2017,samangouei2018defense}), and various types of AE detectors that try to differentiate legitimate inputs and AEs according to specific criteria (e.g.,~\cite{Carrara2018AdversarialED,Shan2020GottaCA}). While effectively improving the robustness of DNN models, to the best of our knowledge, they all suffer from some weaknesses, e.g., defending against only a subset of AEs or causing a relatively high accuracy loss for legitimate inputs. 

More importantly, almost all defense solutions suffer from significant performance degradation with \emph{adaptive attacks}, wherein attackers craft new AEs with the knowledge about the defense solutions. For example, AutoAttack~\cite{croce2020reliable} reports lower robust accuracy on most previous defense solutions by automatically tuning the attack hyperparameters. Recently, the Orthogonal Project Gradient Descent (Orthogonal-PGD) attack~\cite{bryniarski2021evading} has reduced the robust accuracy of several earlier unbroken defenses to nearly zero in the worse case. 

Generally speaking, a DNN model performs complicated non-linear mappings from the high-dimensional input space to a low-dimensional discriminative feature space~\cite{buckman2018thermometer}.
Since the training is conducted on limited samples, it is likely to have loopholes in the model where adjacent inputs differ significantly in the feature space during the mapping. 
We argue that existing AE defense solutions fail to defend against adaptive AEs because \emph{it is extremely challenging, if not impossible, to eliminate or model all the loopholes at the feature space.}

Unlike existing solutions, we do not intend to eliminate or model the loopholes of the DNN model. The \textit{very nature} of AEs is the contradiction between the semantics of the input samples and the model outputs. We propose a novel AE detection framework by directly modeling such contradiction, namely \textit{\contranet}. As we \emph{conduct AE detection at the input space} with \contranet, it is less likely to evade by adaptive attacks.

To be specific, we first train a generator by encoding the legitimate samples and their corresponding labels as the generator's inputs.
The generated synthetic results strive to preserve the semantics of the samples according to their labels. During inference, we feed both the input and the output of the targeted DNN model to obtain the synthetic result.
For legitimate inputs that are correctly inferred, the synthetic output tries to reconstruct the input.
For AEs, instead of reconstructing the input, the synthetic result would be created to conform to the wrong label whenever possible.
Consequently, by estimating the similarity between the input and the synthetic result, we can differentiate AEs from legitimate inputs. 

The contributions of this work include:
\noindent

\begin{itemize}[leftmargin=*, itemsep=3pt]
\item  To the best of our knowledge, \contranet is the first work to explore the AE's intrinsic property for detection at input space, which has the potential to resist adaptive attacks.
\item To realize the potential of \contranet, we propose to generate synthetic samples that keeps the semantics of the inference result and compare with input samples for AE detection. This is achieved by designing a new conditional Generative Adversarial Network (cGAN) network.

\item We develop an effective and efficient similarity measurement model to tell the semantic difference between the input and the synthetic samples.

\end{itemize}

We perform comprehensive evaluations on several popular image classification datasets under various AE attack scenarios. Experimental results show that \contranet outperforms existing solutions by a large margin, especially under adaptive attacks. Moreover, our analysis shows that those successful AEs that can bypass \contranet tend to have much-weakened adversarial semantics. We have also shown that \contranet can be easily combined with adversarial training techniques to further improve the AE defense capabilities of DNN models.   

The remainder of this paper is organized as follows.
Sec.~\ref{sec:background} discusses related work in AE attacks and defenses. Next, we give an overview of \contranet in Sec.~\ref{sec:ContraNet_method}, followed by the concrete design shown in Sec.~\ref{sec:ContraNet_implementation}. Then, we empirically evaluate the performance of \contranet against both white-box attacks and adaptive attacks in Sec.~\ref{evaluation} and Sec.~\ref{sec:Adaptive}, respectively. Sec.~\ref{sec:limitation} discusses the limitations of \contranet. Finally, Sec.~\ref{sec:conclusion} concludes this paper.

%% file: 2.background_relatedworks.tex
\section{Background and Related Works}
\label{sec:background}

\subsection{Adversarial Example Attacks}
\label{sec::AE}

\noindent
\textbf{FGSM.} Goodfellow \textit{et al.}~\cite{goodfellow2014explaining} propose the first simple yet efficient method -- Fast Gradient Sign Method (FGSM) -- to construct AEs against a given DNN classifier. FGSM generates the AE by performing a one-step optimization on the input image towards the gradient ascent direction.

\noindent
\textbf{BIM.} Basic Iterative Method (BIM)\cite{kurakin2016bim} is an iterative variant of FGSM. Under a certain perturbation budget, instead of optimizing the AE in one step as in FGSM, BIM uses smaller steps and iteratively optimizes the AE.

\noindent
\textbf{PGD.} Project Gradient Descent (PGD) proposed by Madry \textit{et al.} \cite{madry-PGD} is a powerful iterative attack method, where the search step starts from a random position in the neighborhood of the clean input.
As PGD relaxes the search direction, it can search AEs with subtle perturbations faster. 
PGD has also been used as a basic building block to construct stronger attacks, e.g., AutoAttack~\cite{croce2020reliable} and Orthogonal-PGD~\cite{bryniarski2021evading}, breaking many state-of-the-art AE defenses.

\noindent
\textbf{C\&W.} 
While the above methods are all variants of FGSM and are based on gradient ascent, C\&W attack~\cite{carlini2018towards} is the first work that formulates the AE generation process as an optimization problem, and it can successfully construct AEs with much smaller perturbations when compared to earlier techniques.

\noindent
\textbf{EAD.}
In~\cite{chen2018ead}, Chen \emph{et al.} treat the adversarial attack as an elastic-net regularized optimization problem and propose the EAD attack. EAD enhances the attack transferability, i.e., the ability of AEs generated against one model being able to attack another unseen model successfully.

\subsection{Defenses against Adversarial Examples}

Existing defense techniques alleviate the impact of AEs by conducting transformation to the inputs, adversarially training the model, or detecting anomalies based on specific criteria.

Gradient masking/ obfuscation schemes (e.g.,~\cite{papernot2017practical, Lcuyer2019CertifiedRT}) try to construct robust models with gradients that are difficult to use by attackers. However, such defense solutions can be easily circumvented with  
black-box attacks such as Zoo~\cite{Chen_2017} or attacks with gradient approximation capabilities~\cite{athalye2018obfuscated}.

Adversarial training methods aim at improving the robustness of a model by adding AEs into the training phase. For example,~\cite{madry-PGD} trains a robust model with AEs generated using PGD attack on the fly.
Other methods include changing the loss function (e.g., TRADE~\cite{zhang2019theoretically}), changing the activation function (e.g.,~\cite{elfwing2018sigmoid}), utilizing artificially generated training examples (e.g.,~\cite{gowal2020uncovering}) or reweighting misclassiﬁed samples (e.g.,~\cite{Wang2020Improving, zhang2021geometryaware}).
There are also adversarial training methods that utilize ensemble learning~\cite{pmlr-v97-pang19a}, metric learning~\cite{li2019improving}, and self-supervised learning~\cite{chen2020adversarial} techniques.
Adversarial training methods are easy to implement, but they inevitably decrease the accuracy of legitimate inputs~\cite{croce2020reliable}.

Input transformation-based defense techniques (e.g.,~\cite{MengMagNetTwoProngedDefense2017,samangouei2018defense}) try to ``purify'' the inputs before feeding them into the DNN model. By doing so, carefully crafted adversarial perturbations are changed, thereby mitigating their attack abilities.  
Due to the nature of this technique, there is an inherent tradeoff between tolerable perturbations and prediction accuracy for legitimate inputs. 
\begin{figure*}[htb]
    \centering
    \includegraphics[width=0.75\linewidth]{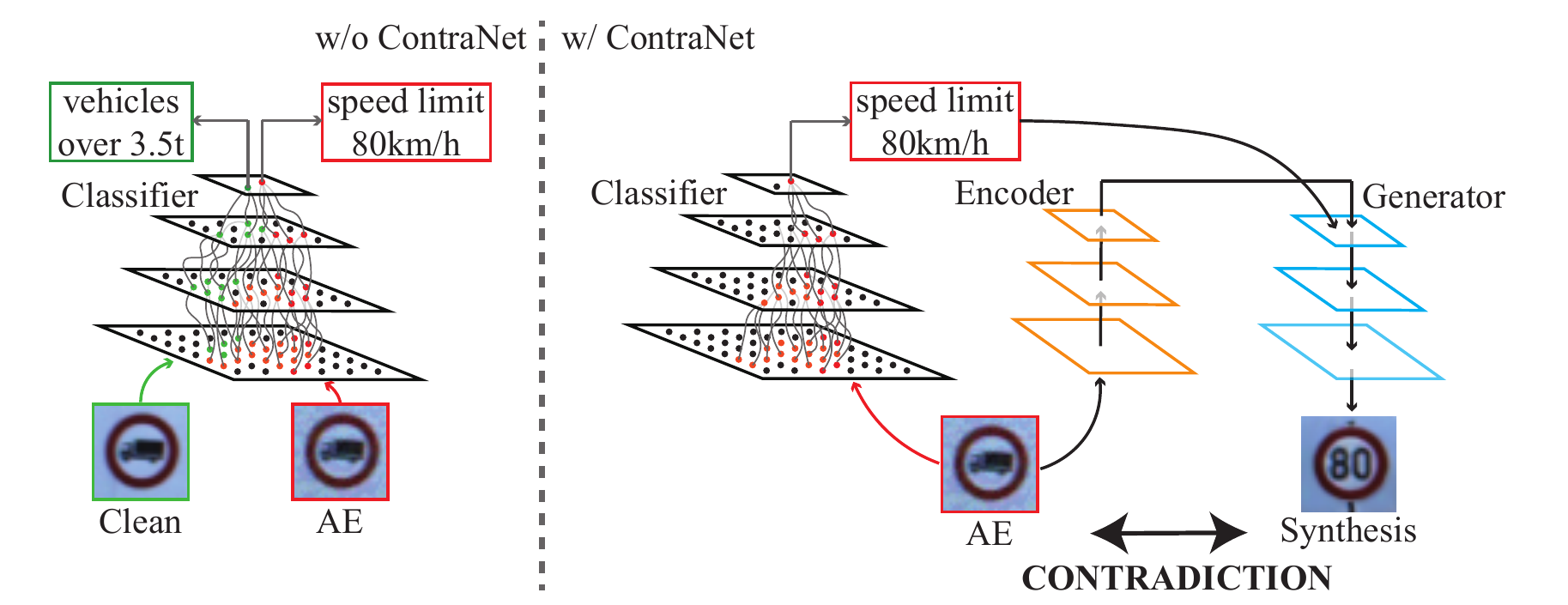}
    \caption{
    The overview of the proposed \textit{ContraNet} solution. The clean input and the crafted AE are similar to each other while getting distinct discriminative features during DNN inference. ContraNet highlights this contradiction by projecting the DNN-extracted discriminative features back to the input space, i.e., ContraNet flags it to be an AE if the synthetic image is dissimilar to the input.
    \vspace{-10 pt}}
    \label{fig:theroy}
   
\end{figure*}

Detection-based methods aim at building an auxiliary module, i.e. an \textit{AE detector}, to reject AEs. Some AE detectors are constructed without changing the target DNN model. For example, 
\textit{MagNet}~\cite{MengMagNetTwoProngedDefense2017} proposes two AE detectors based on reconstruction error and probability divergence, respectively. 
\textit{Feature Squeezing (FS)}~\cite{XuFeatureSqueezingDetecting2018} builds a detector based on the instability of the output from AEs under 
different feature squeezing methods.
Other AE detectors are tightly coupled with the protected DNN model, thus retraining is needed.
In~\cite{Shan2020GottaCA}, Shan \emph{et al.} introduce trapdoors into the target classifier and leads attacks to gravitate towards trapdoors. AEs are identified by comparing neuron activation signatures of inputs to those of trapdoors.
Pang \emph{et al.}~\cite{pang2021adversarial} propose to use true conﬁdence as an indicator of the uncertainty of the predictive results. Since true conﬁdence is not available during inference, they use an MLP branch to learn the rectified conﬁdence as a substitute and reject those samples with low uncertainty as AEs.

As discussed earlier, the existence of AEs is due to the loopholes in the DNN model, which map some adjacent inputs to distant discriminative features. From this perspective, adversarial training techniques try to reduce the number of loopholes with additional samples. However, the added training samples could be far from complete to cover all possible loopholes. Input-transformation defense tries to shift the AE that maps to a loophole point to a safe zone, but its success is not guaranteed, and it may also shift a legitimate input and mistakenly map it to a loophole point. AE detectors try to find one or several criteria to model the loopholes.
Their capability against AE attacks is thus dependent on how easy to bypass the corresponding criteria.

%% file: 3.ContraNet_framework.tex
\section{ContraNet Overview}
\label{sec:ContraNet_method}
\subsection{Threat Model}
\noindent \textbf{Adversarial goals.} Adversarial attacks are divided into two categories: targeted and untargeted attacks.
Targeted attacks succeed when the DNN model outputs a specified incorrect label set by the attacker. In comparison, untargeted attacks lead DNN to provide an arbitrary label that differs from the ground truth~\cite{papernot2016limitations,carlini2018towards}. In this paper, we conduct both targeted and untargeted attacks for the evaluations.

\noindent \textbf{Adversary's Knowledge.}  We consider the two most challenging cases for the defender: white-box and adaptive attacks, as shown in Tab.~\ref{tab:typical_attack_scenario}.
The adversary in a white-box attack scenario has complete knowledge of the classifier, including its network architecture, exact parameters, but the proposed defense is kept secret, i.e. in the white-box attack scenario the attacker is unaware of the existence of the defense method.
Further, in the adaptive attacks\footnote{This is a white-box attack setting for the proposed defense method.}, we assume that the adversary has full access to the proposed defense, i.e., \contranet, including the detection algorithm, the Encoder $\mathbf{E}$, the Generator $\mathbf{G}$, and the Similarity Measurement Model, but not the discriminators, i.e., $D_{\Phi}$ and $D_{aux}$, which are only used in training~\footnote{E, G,  $D_{\Phi}$,  $D_{aux}$ are the components of \contranet whose detailed design information can be found in the following sections}. This adaptive attack scenario represents the most potent adversaries; as informed in ~\cite{CarliniEvaluatingAdversarialRobustness2019, bryniarski2021evading}, a substantial number of existing defenses are broken under adaptive attacks.   

\noindent \textbf{Adversarial Capabilities.}  We impose several reasonable constraints on the adversary.
The adversary cannot destroy the integrity of the model, i.e., the attacker cannot attack the training process, such as poisoning attacks~\cite{steinhardt2017certified,pmlr-v37-xiao15}, backdoor attacks~\cite{lin2020composite,wang2019neural}, nor directly modify the parameters inside the models, e.g., bit-flips attack~\cite{2020,DeepHammer,he2020defending}, fault injection attack~\cite{8203770, breier2018practical}.
Additionally, we assume that the data pre-processing stage cannot be tampered with by an adversary.\footnote{We further discuss the possibilities for implementing \contranet to be resistant to the data pre-processing attack in Sec.~\ref{sec:limitation}.}

\subsection{Design Motivation}

\label{sec::contradiction}
According to the definition of AE~\cite{szegedy2013intriguing}, we deduce two objectives an AE needed to fulfill simultaneously: 1) causing DNN to make a wrong prediction; 2) being indistinguishable from its source image.
The first item ensures the impaction of AE, while the second objective avoids AE being easily identified.
Formally, we summarize the intrinsic feature of AEs as the following corollary: 
\begin{corollary}
    AE's visual semantic information contradicts its discriminative features extracted by the DNN under adversarial attack. 
    \label{cor:feature}
\end{corollary}
Fig.~\ref{fig:theroy}'s left part illustrates this corollary: the clean input and the crafted AE appear almost identical to humans.
However, the DNN model is activated quite differently in the feature space, thus predicting the clean sample as the ``vehicles over 3.5t'' sign and the AE as the ``speed limit 80km/h'' sign.

DNN-based image classifiers perform complicated non-linear mappings from the high-dimensional input image space to the low-dimensional feature space.
Especially considering that DNN models are constructed with limited training samples, the multi-to-one mapping from DNN models leaves many loopholes for attackers to generate possible AEs that are similar to the clean sample in the image space but differ significantly in the feature space.

Directly checking whether the semantic consistency holds between the input image space and the feature space is intractable.
However, if we project the inference result of the DNN model back to the input space, the contradiction within Corollary.~\ref{cor:feature} can be reflected in the image space.
We could then compare it with the original input at the same space, 
as shown in the right part of Fig.~\ref{fig:theroy}.
This is the key idea of ContraNet.

\subsection{Overview of ContraNet}
\label{sec::overview of ContraNet}
To perform semantic comparison at the input space, ContraNet employs an Encoder to capture the low-level features of the input image, and feed it together with the inference result of the DNN model to a Generator. The semantics of the  synthetic image would conform to the DNN model's output,
e.g., the semantics of the synthesis shown in the right part of Fig.~\ref{fig:theroy} is faithful to DNN's prediction, i.e., ``speed limit 80 km/h''.
We can then easily tell the difference between the AE input and the synthetic image and use it for AE detection.

\begin{figure}[t]
    \centering
    \includegraphics[width=0.98\linewidth]{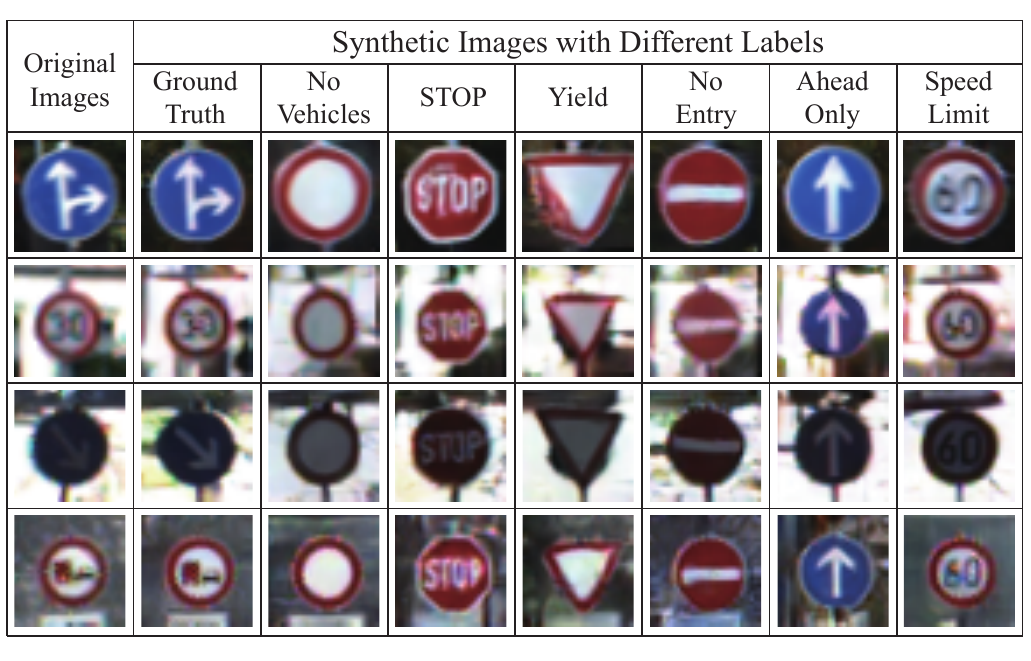}
    \caption{The semantic meaning of the synthesis $x'$ is faithful to  the conditional label $y$. In each row, the first element is the input image $x$, and the second element depicts the generated synthetic images conditioned on the ground truth and other labels.   
    }
    \vspace{-5pt}
    \label{fig:project_label_to_inputs}
\end{figure}

To further demonstrate our motivation, we empirically show some cases in Fig.~\ref{fig:project_label_to_inputs}, where \textit{the synthetic images' semantic information is highly dependent on the inferred result from the classifier}.
To be specific, given input images in the first column, we generate synthetic images (in other columns) by varying the label (shown on the top) associated with the generation.
Such procedures mimic the semantics changes caused by AEs at the feature space.
As can be observed, ContraNet faithfully reconstructs the synthetic images when the label is congruent with the input's semantics. However, when the given label contradicts the input image's semantics, the generated synthetic image would be pretty different from the input image. Due to space limitations, more empirically generated syntheses are provided in Appendix~\ref{appendix:more_synthesis}, as shown in Fig.~\ref{fig:more_fig}.

\subsection{Resistance to Adaptive Attacks}

Because ContraNet is designed based on the semantic contradiction, attackers cannot inject imperceptible noise-like perturbations (without any semantic information) to bypass ContraNet. Instead, the adversarial perturbations need to be added toward the target class. However, such kinds of perturbations naturally weaken the AEs' evil properties, as they would cease to be malevolent if the semantics of the perturbed image, follows that of the DNN output. In other words, the only way to eliminate the semantic contradiction is to perturb the image to be alike with a clean image from another class. Such perturbations are not malicious anymore.

%% file: 4.one_implementation.tex
\section{Concrete ContraNet Design}
\label{sec:ContraNet_implementation}

In this section, we detail the proposed \contranet design. 
First, we depict an overview of \contranet's implementation. Then, we discuss several key adaptations for cGAN that have aided the implementation of \contranet's core concept.
To further improve \contranet's detection capability, we propose a similarity measurement model for determining the similarity between the input image and its synthesis.
Finally, we provide a training process for \contranet.

\begin{figure}[t]
    \centering
    \includegraphics[width=\linewidth]{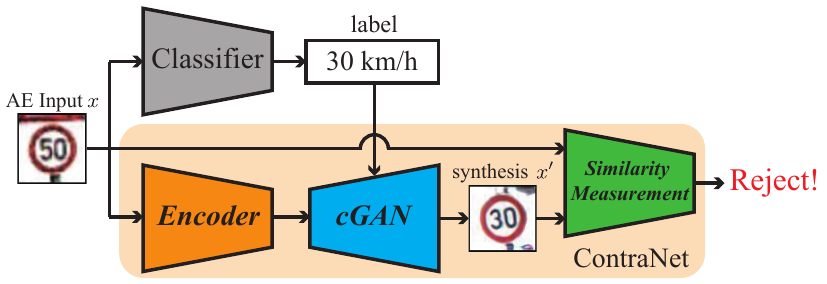}
    \caption{The proposed ContraNet design, containing an Encoder, a conditional Generator (cGAN), and a Similarity measurement model.}
    \vspace{-15pt}
    \label{fig:overview}
    
\end{figure}

\subsection{Implementation Overview}
As depicted in Fig.~\ref{fig:overview}, \contranet consists of three components, an Encoder ($\mathbf{E}$), a Generator ($\mathbf{G}$), and a Similarity measurement model.
To be more specific, the encoder $\mathbf{E}$ is in charge of extracting the low-level features of the input image $x$. The generator $\mathbf{G}$  is built based on a \textit{class-conditional GAN} (cGAN)~\cite{MiyatoCGANsProjectionDiscriminator2018, brock2018large}. Specifically, $\mathbf{G}$ takes as input the low-level features of $x$ summarized by $\mathbf{E}$, while the predicted label of $x$ given by the classifier serves as the conditional input.
The main purpose of $\mathbf{G}$ is to synthesize an image $x'$ whose low-level features such as colors and textures are faithful to the input image $x$, while its semantics conform to the conditional input, i.e., the predicted label of $x$.
In this way, the semantic meaning of the synthesized image $x'$ is highly related to the label given by the classifier.
The Similarity measurement model measures the similarity between the input image $x$ and its synthesized counterpart $x'$.


During inference, given an input $x \in \mathcal{X}_{\text{test}}$, we first generate its synthetic counterpart $x'$ with $x'=\mathbf{G}(\mathbf{E}(x), y)$, conditioned on its predicted label $y = C(x)$, where $C(\cdot)$ indicates the classifier.
Specifically, for a legitimate input that is correctly inferred, its DNN-extracted feature is consistent with its semantic information.
Therefore, its synthetic image $x'=\mathbf{G}(\mathbf{E}(x), y)$ would be similar to itself $x$.
Otherwise, 
the synthetic image's semantic information will be substituted to the predicted label $y$.
Lastly, the similarity measurement model will evaluate the similarity between $x$ and $x'$, to identify AEs.

\begin{figure}[t]
    \centering
    \includegraphics[width=\linewidth]{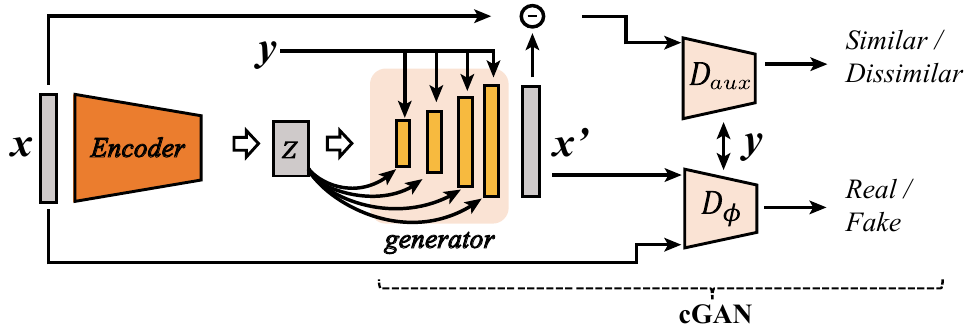}
    \caption{The revised cGAN in ContraNet. The Encoder summaries $x$'s low-level features as $z$, and splits $z$ into several slices to impact generator's deeply. $D_{\Phi}$ is cGAN's vanilla component, while $D_{aux}$ is specific for our task, used to evaluate similarity.}
    \vspace{-15pt}
   
    \label{fig:cvae}
\end{figure}

\subsection{Encoder and cGAN in Cooperation}
The Encoder, \E, and the cGAN's generator, \G, are responsible for projecting the predicted label $y$ back to image space and obtaining a synthetic image whose low-level features are faithful to the input image, and its semantic meaning highly depends on the given $y$.
In this section, we introduce several critical design considerations when implementing these two models.
The overview of Encoder and cGAN is in Fig.~\ref{fig:cvae}.

\noindent
\textbf{Enhancing $y$'s influence.}
As discussed in Sec.~\ref{sec:ContraNet_method}, to realize the detection mechanism of \contranet, the synthetic image's semantic information should significantly correlate to the classifier's predicted label, $y$.
To enhance the influence of $y$ on the synthetic image's semantic meaning, we adopt the Conditional Batch Normalization (CBN) ~\cite{dumoulin2016learned,de2017modulating} technique on cGAN's generator. Within CBN, $y$ directly influences each batch normalization layer's mean and variance parameters. Therefore, the $y$'s impaction can spread to the generator's deeper layers than merely concatenate $y$ to the input layer.
For the discriminator part, we employ the conditional projection layer technique ~\cite{MiyatoCGANsProjectionDiscriminator2018}. The projection layer enhances $y$'s influence by mapping the one-hot label $y$ to a high-dimensional embedding and projecting the embedding onto cGAN's discriminator's middle layer. This projection structure enables high-quality class information transformation. 

\noindent
\textbf{Getting a meaningful latent vector.}
One requirement of \contranet is that the synthesis $x'$ should be faithful to $x$. 
In this case, given a consistent pair of $x$ and $y$, the synthetic image $x'$ should be similar to $x$.
To this end,
we develop an Encoder to summarize the input's low-level features to the latent vector $z=\mathbf{E}(x)$, as shown in Fig.~\ref{fig:cvae}.
We employ $\mathcal{SSIM}$ and ${\ell}_2$ between the input image and its synthesis as constraints.
We also apply a \textit{KL-Divergence} loss, $\mathcal{L}_{D_{KL}}$, to regular $z\sim\mathcal{N}(0, \mathbf{I})$~\cite{doersch2016tutorial}, which guarantees the latent vector to be representative enough. 

\noindent
\textbf{Let the latent vector $z$ impact the deeper layers of cGAN's generator.}
As introduced above, we encode $x$'s low-level features as a latent vector $z$ with \E, then \G is trained to synthesize $x'$, which resembles $x$ given a consistent $y$.
In order to make better use of the information provided by $z$, following~\cite{brock2018large},  
we chunk the $z$ into several slices, $z_{i}$, and distribute these $z_{i}$ into deeper layers of the generator, as shown in Fig.~\ref{fig:cvae}'s \textit{generator} part. 
In this way, the influence of $z$ can be extended to deeper layers.  The synthetic image $x'$ can reconstruct $x$ as much as possible when the given label consistent with $x$'s original semantic meaning.

\noindent
\textbf{Coupling an auxiliary discriminator.}
\label{sec::discriminator2}
In the vanilla cGAN's adversarial training process, the discriminator, designated by $D_{\Phi}$, is responsible for distinguishing generated images from real ones. 
While the generator is optimized to generate realistic images that trick the discriminator, $D_{\Phi}$ since the original task of vanilla cGAN is to generate realistic and various images.
Whereas in our task, an additional criterion must be met, i.e., the synthetic image, $x'$, should be faithful to its corresponding input image, $x$, given a consistent, $y$.
As a result, adding an auxiliary discriminator, $D_{aux}$, is appropriage to determine if the synthesized image $x'$ is close to the corresponding input $x$. 

To be specific, the auxiliary discriminator, $D_{aux}$, focuses on the difference between the input  and its synthesis.
We use $x - x'$ as the input directly.
The positive input, $x_{pos}$, is defined as the difference between $x$ and its synthetic image under a consistent label $y$,  denoted as $x'_{y}$ (Eq.~\eqref{eq:dis2pos}).
The negative input, $x_{neg}$, is defined as  $x$ minus its synthetic image generated under an inconsistent label $y'$, denoted as $x'_{y'}$ (Eq.~\eqref{eq:dis2neg}). 
\vspace{-17pt}

\begin{small}
\begin{align}
    x_{pos} &= x - x'_{y} \label{eq:dis2pos}\\
    x_{neg} &= x - x'_{y'} \label{eq:dis2neg}
\end{align}
\end{small}
\vspace{-17pt}

\noindent
We employ the \textit{hinge loss} (Eq.~\eqref{equ:hingeloss}) as the criterion for $D_{aux}$, which will push the distance of positive input to be larger than $1$, while pulling that of negative input to be less than $-1$. 
\vspace{-17pt}

\begin{small}
\begin{multline}
    \label{equ:hingeloss}
    \mathcal{L}_{D_{aux}} = ReLU(1-D_{aux}(x_{pos}, y)) \\+ ReLU(1+D_{aux}(x_{neg}, y))
\end{multline}
\end{small}
\vspace{-25pt}

\subsection{Similarity measurement model}
\label{sec::similarity_measurement}

We devise a similarity measurement model to distinguish the similarity between $x$ and $x'$. If $x$ and $x'$ are deemed dissimilar, \contranet will reject $x$; otherwise, $x$ is believed to be a benign sample.
We make the judgment from three different perspectives.
The proposed Deep Metric Model (\dmm) provides similarity judgment by comparing a hierarchical distance between $x$ and $x'$.
In contrast $\ssim$ provides a pixel-to-pixel similarity judgment.
Further, we apply a \dis model to reject a sample according to the quality of its synthesis. 
The insight of \dis is that if the given label does not match a sample's semantic meaning, the synthesis will appear more unrealistic than the synthesis generated with the accordant label.
These three rejectors work in a tandem way, where any rejection of the three leads to a final rejection.

\noindent
\textbf{Deep Metric Model.} The insight of deep metric learning is to develop a network that distinguishes similar from dissimilar example pairs by learning a hierarchical distance. 
Deep metric learning, e.g., siamese network~\cite{koch2015siamese} and triplet network~\cite{hoffer2018deep}, is widely used for similarity measurement tasks, such as image retrieval, face recognition and signature verification~\cite{bromley1993signature,schroff2015facenet,wang2014learning}.
Inspired by the above concept, we adopt a triplet network to our \dmm by labeling images from the same class as positive pairs while images from different classes as negative pairs.
Since there are slight differences on the data distribution between $x$ and $x'$, we use two models, $\mathbf{F_{real}}$ and $\mathbf{F_{syn}}$ trained on $x$ and $x'$, respectively, as their feature extractors.


After getting the two pretrained feature extractors,
we treat the concatenated embeddings {\small{$ep = Concat(\mathbf{F_{real}}(x), \mathbf{F_{syn}}(x'_y))$}} as positive while
{\small{$en = Concat(\mathbf{F_{real}}(x), \mathbf{F_{syn}}(x'_{y'}))$}} as negative. 
Finally, we train a Multi-Layer Perceptron (MLP) model with $en$, $ep$ as input and the final similarity decision as its output.
We construct this MLP by three fully-connected layers.
We use \textit{Cross-Entropy loss}, denoted by $\mathcal{L}_{CE}$~\cite{board2005stochastic}, as the objection function for this binary classification task.
In our experiments, we incorporate AEs generated by PGD attack~\cite{madry-PGD} into the training set to improve the robostness of the \dmm.
A more detailed training process is provided in Algorithm~\ref{alg:training_framework}, line 22-29.

\noindent
{\boldmath{$\ssim$}}. $\ssim$ is a typical similarity metric. $\ssim$ judges the similarity between $x$ and $x'$ in pixel-level without the learning process. Thus, $\ssim$ can be  used directly as a rejector, which flags AEs with low $\ssim$ similarity.


\noindent
\textbf{Discriminator.}
During the training process of the generator, since the input label is accordant with the input image, the synthetic image $x'$ will be an exact reconstruction of the input image $x$, and appear realistic. 
However, when $y$ contradicts $x$, $x'$ could be regarded as an image formed by substituting $x$'s semantic information with $y$'s image-space projection.
Because of the contradiction, the generated $x'$ will look unrealistic and be of poor quality. 
Based on the above observation, we propose to use a Discriminator, \dis, as a rejector.
Its objective is set in Eq.~\eqref{eq:Dis_loss}.
Note that \dis receives only the synthetic image $x'$ as input and rejects a sample according to its synthesis's quality. 
It is difficult for the adversary to influence \dis's decision by adding subtle adversarial perturbation to $x$.
Therefore, \dis can promote the robustness of \contranet against AEs.

\vspace{-17pt}
\begin{small}
\begin{multline}
    \label{eq:Dis_loss}
    \mathcal{L}_{Dis} = ReLU(1-Dis(x_{y}', y))+ ReLU(1+Dis(x_{y'}', y))
\end{multline}
\end{small}
\vspace{-17pt}

\subsection{Implementation details}
The whole \contranet consists of the Encoder $\mathbf{E}$, the Generator $\mathbf{G}$, the two discriminators $D_{\phi}$ and $D_{aux}$, and the similarity measurement model (\dmm and \dis). We train \contranet using Adam~\cite{Bengio2014Adam} with $\beta_{1}=0.9$, $\beta_{2}=0.99$, the batch size setting at 128. We use a typical linear learning rate decay strategy with a starting point at $0.0005$. Following ~\cite{yaz2018unusual}, we employ EMA with decay factor at $0.9999$ as Generator's regularization technique. We adopt adding random noise and \textit{DiffAugment} ~\cite{zhao2020differentiable} as data augmentation. 
The overall training algorithm for \contranet has been summarized as Algorithm~\ref{alg:training_framework}, and we provided the detailed model architectures in Appendix~\ref{appendix:architectures}.

\begin{algorithm}[hbt]
    \caption{\contranet Training Framework}
    \label{alg:training_framework}
    \begin{algorithmic}[1]
        \footnotesize
        \Statex \uit{\textbf{Stage1: train ContraNet's Encoder and Generator}}
        \Statex \textbf{Input:} Training data $\mathcal{X} = \{x\}^N$, $\mathcal{Y}=\{y\}^N$
        \Statex \textbf{Output}: The parameters of $\mathbf{E}$, $\mathbf{G}$, \bm{$D_{\Phi}$}
        \Statex \Comment Train $\mathbf{G}$, $\mathbf{E}$ and \bm{$D_{\Phi}$}.
        \For {some training iterations}
        \State $x'_{y} = \mathbf{G}(\mathbf{E}(x), y)$;
        \State Feed $x$, $x'_{y}$ and $y$ into \bm{$D_{\Phi}$};
        \State Optimize \bm{$\mathbf{G}$} and \bm{$\mathbf{E}$} for $\mathcal{L}_{\mathbf{E}+\mathbf{G}}$ (Eq.~\eqref{eq:cGAN_gen});
        \State Optimize \bm{$D_{\Phi}$} for $\mathcal{L}_{D_{\Phi}}$ (Eq.~\eqref{eq:dis1});
        \EndFor
        \Statex \Comment Train \bm{$D_{aux}$} with fixed $\mathbf{G}$ and $\mathbf{E}$.
        \For {some training iterations}
        \State $x_{pos} = x - \mathbf{G}(\mathbf{E}(x), y)$, $x_{neg} = x - \mathbf{G}(\mathbf{E}(x), y')$;
        \State Feed $x_{pos}$, $x_{neg}$ and $y$ into \bm{$D_{aux}$};
        \State Optimize $\bm{D_{aux}}$ for $\mathcal{L}_{D_{aux}}$ (Eq.\eqref{equ:hingeloss});
        \EndFor
        \Statex \Comment Fine-tune $\mathbf{G}$, $\mathbf{E}$ and \bm{$D_{\Phi}$} with fixed \bm{$D_{aux}$}.
        \For {some training iterations}
        \State $x'_{y} = \mathbf{G}(\mathbf{E}(x), y)$, $x'_{y'} = \mathbf{G}(\mathbf{E}(x), y')$; 
        \State $x_{pos} = x - x'_{y}$, $x_{neg} = x - x'_{y'}$;
        \State Feed $x_{pos}$ and $x_{neg}$ into \bm{$D_{aux}$} to compute $\mathcal{L}'_{\mathbf{E}+\mathbf{G}}$ (Eq.~\eqref{eq:cGANprime});
        \State The same as line 3-5;
        \EndFor\\
    
        \Statex \uit{\textbf{Stage2: train Similarity measurenet model's} \bm{$Dis$}}
        \Statex \textbf{Input:} Training data $\mathcal{X} = \{x\}^N$, $\mathcal{Y}=\{y\}^N$, fixed $\mathbf{E}$ and $\mathbf{G}$. 
        \Statex \textbf{Output}: The parameters of \bm{$Dis$}.
        \For {some training iterations}
        \State $x'_{y} = \mathbf{G}(\mathbf{E}(x), y)$, $x'_{y'} = \mathbf{G}(\mathbf{E}(x), y')$;
        \State Feed $x'_{y}$, $x'_{y'}$ and $y$ into \bm{$Dis$} and optimize for $\mathcal{L}_{Dis}$ (Eq.~\eqref{eq:Dis_loss});
        \EndFor\\
        \Statex \uit{\textbf{Stage3: train Similarity measurenet model's} {$\mathbf{DMM}$}}
        \Statex \textbf{Input:} Training data $\mathcal{X} = \{x\}^N$, $\mathcal{Y}=\{y\}^N$,  fixed $\mathbf{E}$ and $\mathbf{G}$. 
        \Statex \textbf{Output}: The parameters of $\mathbf{DMM}$ including $\mathbf{F_{real}}$, $\mathbf{F_{syn}}$ and $\mathbf{MLP}$.
        \State Train feature extractor $\mathbf{F_{real}}$ on $x$;
        \State Train feature extractor $\mathbf{F_{syn}}$ on $x'=\mathbf{G}(\mathbf{E}(x),\cdot)$;
        \Statex \Comment{Train $\mathbf{DMM}$ with pretrained $\mathbf{F_{real}}$ and $\mathbf{F_{syn}}$}
        \For {some training iterations}
        \State $e_p$ = $\operatorname{Concat}(\mathbf{F_{real}}(x), \mathbf{F_{syn}}(x'_{y}))$, label as positive;
        \State $e_n$ = $\operatorname{Concat}(\mathbf{F_{real}}(x), \mathbf{F_{syn}}(x'_{y'}))$, label as negative;
        \State Feed $e_p$, $e_n$ to $\mathbf{MLP}$
        \State Optimize $\mathbf{F_{real}}$, $\mathbf{F_{syn}}$ and $\mathbf{MLP}$ for $\mathcal{L}_{CE}$;
        \EndFor
\end{algorithmic}
\end{algorithm}

\noindent
\textbf{Training process.} To obtain  \E and \G, there are four models needed to be trained: an encoder ($\mathbf{E}$), a generator ($\mathbf{G}$), a discriminator ($D_{\Phi}$), and an auxiliary discriminator ($D_{aux}$).
The \E, \G, $D_{\Phi}$ together can be viewed as a revised cGAN. Therefore, we train these three models following the standard GAN training process. The adversarial loss for the revised cGAN's generator part (\E and \G) is denoted as $\mathcal{L}_{\mathbf{E}+\mathbf{G}}$ in Eq.~\eqref{eq:cGAN_gen}:

\vspace{-5pt}
\begin{small}
\begin{equation}
    \label{eq:cGAN_gen}
            \mathcal{L}_{\mathbf{E}+\mathbf{G}}=\mathcal{L}_{D_{KL}}+ {\ell}_2 + \mathcal{SSIM} + \mathcal{L}_{\mathbf{G}},
    \end{equation}
\end{small}
\vspace{-10pt}

\noindent where {\small{$\mathcal{L}_{\mathbf{G}}=ReLU(1-D_{\Phi}(x'_{y},y))$}} is the GAN loss on $\mathbf{G}$.
Furthermore, the adversarial loss for the revised cGAN's discriminator part ($D_{\Phi}$) can be written as Eq.~\eqref{eq:dis1}.

\vspace{-17pt}
\begin{small}
\begin{equation}
    \label{eq:dis1}
    \mathcal{L}_{D_{\Phi}} = ReLU(1-D_{\Phi}(x, y)) + ReLU(1+D_{\Phi}(x'_{y}, y))
\end{equation}
\end{small}
\vspace{-17pt}

\noindent
The training procedure of this revised cGAN is demonstrated in Algorithm~\ref{alg:training_framework}, line 1-6.

The purpose of $D_{aux}$ is to improve the similarity between $x$ and $x'$ when the given conditional label matches $x$'s semantic information. Thus, adding $D_{aux}$ in the latter training phase when  \E and \G can already generate meaningful synthesis makes sense. As opposed to letting $D_{aux}$ join the training from scratch, adding $D_{aux}$ in the later training phase reduces the difficulty of optimization and makes the training process smoother and easier to converge.  To be more specific, first, we obtain $D_{aux}$ by training it with fixed \E and \G, which can already generate realistic but similarity unsatisfactory synthesis,
as demonstrated in Algorithm~\ref{alg:training_framework} line 7-11. 
Then we freeze $D_{aux}$ and add it to the $\mathcal{L}_{\mathbf{E}+\mathbf{G}}$ and get the fine-tune loss, $\mathcal{L}'_{\mathbf{E}+\mathbf{G}}$, as demonstrated in Eq.~\eqref{eq:cGANprime}. We show the fine-tuning procedure in Algorithm~\ref{alg:training_framework} line 12-17.

\vspace{-10pt}
\begin{small}
\begin{equation}
\label{eq:cGANprime}
        \mathcal{L}'_{\mathbf{E}+\mathbf{G}}=\mathcal{L}_{D_{KL}}+ {\ell}_2 + \mathcal{SSIM} + \mathcal{L}_{\mathbf{G}}  + \mathcal{L}_{D_{aux}}
\end{equation}
\end{small}
\vspace{-10pt}

\noindent
Note that $D_{\phi}$ and $D_{aux}$ are only auxiliary models to help the revised cGAN consider the label information. Only \E and \G are used during inference.

%% file: 6.whitebox_attack_evaluation.tex
\section{Evaluation}

\label{evaluation}
This section reports \contranet's performance against white-box attacks. Following previous defenses~\cite{Shan2020GottaCA,pang2021adversarial,MengMagNetTwoProngedDefense2017,XuFeatureSqueezingDetecting2018}, we evaluate \contranet on four popular adversarial attacks together with a new adversarial attack benchmark, \textit{AutoAttack}.~\cite{croce2020reliable} 
Note that a more rigorous evaluation against adaptive attacks is given in Sec.~\ref{sec:Adaptive}.  

\vspace{-5pt}
\subsection{Experimental Settings}
\label{sec::exp_settings}
\noindent
\textbf{White-box attack.} In the white-box attack setting, the attacker has complete knowledge of the classifier, whereas the detector is confidential.

\noindent
\textbf{Datasets.}
We conduct experiments on \mnist, \gtsrb, and \cifar, which are the
\emph{de facto} datasets used to evaluate AE defenses.
We use commonly adopted classifier architectures, such as ResNet18, DenseNet169. 
More details of the datasets and classifiers are in Tab.~\ref{tab:classifier}.

\begin{table}[t]
	\centering
	\caption{Information of Datasets and Classifiers.}
	\label{tab:classifier}
	\begin{tabular}{@{}lccclc@{}}
	\toprule
	Dataset & \# of Class & Trainset Size & Testset Size & \begin{tabular}[c]{@{}l@{}}Classifier Architecture\end{tabular} \\
	\hline
	\mnist~\cite{726791}   & 10 & 50,000 & 10,000 & 2Conv, 2FC \\
	\gtsrb~\cite{Stallkamp2012}   & 43 & 35,288 & 12,630 & ResNet18\cite{he2016deep} \\
	\cifar~\cite{krizhevsky2009learning} & 10 & 50,000 & 10,000 & DenseNet169\cite{huang2017densely} \\
	\bottomrule
	\end{tabular}
	\vspace{-10pt}
\end{table}

\noindent
\textbf{Evaluation metrics.}
The metrics employed evaluate ContraNet from two perspectives: 1) the impact on classification accuracy on legitimate samples, and 2) the ability of AE detection.
We conventionally denote the adversarial example as the {\it Positive sample} (P) and the clean sample as the {\it Negative sample} (N).
Unless specified, we use the same evaluation metrics for the adaptive attack in Sec.~\ref{sec:Adaptive}.
\noindent

\begin{itemize}[leftmargin=*, itemsep=3pt]
	\item \textbf{Detector's metric.} TPR@FPR n\% 
	indicates the True Positive Rate (TPR) when fixing the threshold with False Positive Rate (FPR) $\le n\%$.
	This metric evaluates the performance of the detector alone.
	The FPR describes the fraction of normal samples being flagged as AE.
	In general, lower FPR induces lower TPR.
	Therefore, there is a trade-off between TPR and FPR depending on the application scenario.
	Commonly, TPR@FPR 5\% serves as the primary metric to evaluate the detector's performance~\cite{Pang2020ATO,Shan2020GottaCA,MengMagNetTwoProngedDefense2017}. In this work, we report the typical 
	TPR@FPR 5\% as the main indicator. The TPR at a lower FPR, TPR@FPR3\%, is also provided to demonstrate \contranet's performance further. 

	\item \textbf{Detector's accuracy on clean samples.} $Acc_{dec}$
	indicates the detector's accuracy on clean samples by combining the detector with the classifier, as addressed in Eq.~\eqref{eq:clean_acc}.
	This metric reflects ContraNet's impact on clean samples.

	\begin{footnotesize}
	\begin{multline}
		\label{eq:clean_acc}
		Acc_{dec} =  \frac{\#Classifer\ correct\&Detector\ pass}{\#all\ clean\ samples} + \\ \frac{\#Classifier\ wrong\&Detector\ reject}{\# all\ clean\ samples}
	\end{multline}
	\end{footnotesize}
	 
	\item \textbf{ROC curve \& AUC}. We plot Receiver Operating Characteristic (ROC) curves to elaborate the influence of the various threshold settings.
	A more general metric, Area Under the Curve (AUC), is provided to give an overall summary for its corresponding ROC curve.

	\item \textbf{Robust accuracy on AEs.} $Acc_{rob}$ demonstrates the attacker's failure rate, as  shown in Eq.~\eqref{eq:robust_acc}, where the {\it Successful AEs} are AEs that fool the classifier and bypass the detector, and {\it all AEs} are samples that have been perturbed by attacks.
	This metric follows \cite{MengMagNetTwoProngedDefense2017, Shan2020GottaCA,Pang2020ATO,bryniarski2021evading}.
	$Acc_{rob}$ can well reflect the overall performance of the whole system (considering both the classifier and the detector).
	
	\begin{footnotesize}
	\begin{equation}
		\label{eq:robust_acc}
		Acc_{rob} = 1 - \frac{\#Successful\ AEs}{\#all\ AEs}
	\end{equation}
	\end{footnotesize}
\end{itemize}

\noindent
\textbf{Baselines.} We choose four detection-based defense methods as the baselines, including two high cited schemes: MagNet~\cite{MengMagNetTwoProngedDefense2017} and Feature Squeezing (FS)~\cite{XuFeatureSqueezingDetecting2018} and two most recent schemes: Trapdoor~\cite{Shan2020GottaCA} and Rectified Rejection (RR)~\cite{pang2021adversarial}.
Although most baselines have been bypassed under adaptive attacks by later literature or the authors themselves~\cite{he2021feature, HeAdversarialExampleDefenses2017,carlini2017magnet}, they still get excellent performance against attackers with limited power, e.g., the white-box attacks.
Thanks to their authors, we reproduce their work with officially open-source codes (For TensorFlow projects, we re-implement them using Pytorch).

\subsection{ContraNet against White-box Attacks}\label{sec:ContraNet against white-box attacks}
We evaluate the performance of ContraNet across three datasets with their associated classifiers, as shown in Tab.~\ref{tab:classifier}.
Note that, Trapdoor and RR require specially/adversarially trained classifiers. We show respect to this setting when implementing their methods.
We use four typical untargeted attacks for evaluation, including the iterative attack PGD\cite{madry-PGD} and BIM\cite{kurakin2016bim}, and optimization-based C\&W\cite{carlini2018towards} and EAD\cite{chen2018ead}.
These attacks are broadly used as evaluations of our baselines. We included the exact attack parameter settings in Appendix~\ref{sec::attack_Configuration}.
The misclassification rate can achieve over 98\% for every attack on the vanilla classifier.

\begin{table}[t]
	\caption{Detector's Accuracy on Clean Sample ($Acc_{dec} \uparrow$)}
	\label{tab:normal accuracy}
	\centering
	\begin{threeparttable}
	\setlength{\tabcolsep}{1.5mm}{
	\begin{tabular}{@{}llcccccc@{}}
	\toprule
	Dataset & $Acc_{ori}$\tnote{*} &\textbf{ContraNet} & Trapdoor & RR 		 & MagNet      & FS      \\
	\midrule
	\cifar~\cite{krizhevsky2009learning} & 95.50\% &\textbf{92.02\%}   & 80.22\%  & 82.69\%   & 88.30\% & \uit{91.38\%} \\
	\gtsrb~\cite{Stallkamp2012}   & 98.90\% &\textbf{95.04\%}   & 92.11\%  & 92.09\%   & 91.03\% & \uit{94.12\%} \\
	\mnist~\cite{726791}   & 99.8\%  &\uit{95.56\%}      & 94.40\%  & \textbf{95.75\%}   & 90.13\% & 94.2\%  \\
	\bottomrule
	\end{tabular}}
	\begin{tablenotes}
		\footnotesize
		\item[*] indicates the classification accuracy on clean samples without detectors.
		\item[$\diamond$] The \textbf{bolded} values are the highest performance. The \uit{underlined italicized} values are the second highest performance.  
	\end{tablenotes}
	\end{threeparttable}
\vspace{-10pt}
\end{table}

\noindent\textbf{Impact on accuracy with clean images.}
Tab.~\ref{tab:normal accuracy} summarizes the performance of ContraNet and baselines on clean samples after fixing the threshold @FPR5\%. As can be seen, the accuracy decrease caused by \contranet is at most 4.5\% across three datasets, while that of RR is 12.8\%, Trapdoor 15.3\%, MagNet 9.7\%, FS 5.6\%.
ContraNet causes little impact on the original accuracy, especially when compared to methods requiring re-training the classifier, e.g., RR and Trapdoor.

\noindent
\textbf{Robust Accuracy with Detector.}\label{sec:acc_rob whitebox}
\contranet maintains a high $Acc_{rob}$ across a variety of attacks and datasets, as seen in Tab.~\ref{tab:robustacc}.
The above attack-agnostic property is thanks to ContraNet's detection mechanism.
By using the semantic contradiction, \contranet identifies AEs without the need for classifier knowledge or assuming the attack types.
Consequently, \contranet becomes more general than non-semantic-based methods.
For more results under ${\ell}_{2}$ attack, please refer to Appendix~\ref{appendix:l2_robust_acc}.

\begin{table*}[ht]
	\caption{$Acc_{rob}$ under White-box Attack}
	\label{tab:robustacc}
	\centering
\begin{threeparttable}
	\setlength{\tabcolsep}{5mm}{
	\begin{tabular}{@{}llcccccc@{}}
	\hlinew{1pt}
	\multicolumn{2}{c}{Dataset}   & \cifar         & \multicolumn{1}{c}{\gtsrb} & \multicolumn{1}{c}{\mnist} & \cifar        & \multicolumn{1}{c}{\gtsrb} & \multicolumn{1}{c}{\mnist}\\ \hline
	Attack Method & Defense            & \multicolumn{3}{c}{\bm{$Acc_{rob}$}\textbf{@FPR5\%}$\uparrow$  }   & \multicolumn{3}{c}{\bm{$Acc_{rob}$}\textbf{@FPR3\%}$\uparrow$  }       \\ \hline
				  
	\multirow{5}{*}{$PGD_{{\ell}_{\infty}}$} 
				  & \textbf{ContraNet} & \textbf{90.58\%} & \textbf{97.63\%} & \textbf{100\%}   & \textbf{81.59\%}       & \textbf{95.05\%}          & \textbf{100\%} \\
				  & Trapdoor           & 58.55\%          & \uit{94.95\%}    & 99.88\%   		& 55.59\%                & {\uit{94.75\%}}           & 99.88\%        \\
				  & RR                 & 57.15\%          & 75.86\%          & 99.34\%  		& 54.73\%                & 74.01\%                   & 98.97\%         \\
				  & MagNet             & \uit{74.12\%}    & 47.50\%          & \textbf{100\%}   & {\uit{72.56\%}}        & 43.09\%                   & \textbf{100\%}  \\
				  & FS                 & 63.57\%          & 91.47\%          & 96.69\%          & 56.05\%                & 85.98\%                   & 96.61\%      \\
	\hline
	\multirow{5}{*}{$BIM_{{\ell}_{\infty}}$} 
				  & \textbf{ContraNet} & \textbf{91.45\%} & \uit{97.55\%}    & \textbf{100\%}   & \textbf{81.29\%}       & {\uit{94.30\%}}           & \textbf{100\%}  \\
				  & Trapdoor           & 58.55\%          & \textbf{97.78\%} & 99.51\%          & {\uit{72.08\%}}        & \textbf{97.16\%}          & \uit{99.47\%}        \\
				  & RR                 & 54.24\%          & 71.75\%          & 99.3\%           & 51.85\%                 & 70.39\%                   & 98.97\%      \\
				  & MagNet             & \uit{61.35\%}    & 44.51\%          & \textbf{100\%}   & 59.86\%                & 40.39\%                   & \textbf{100\%} \\
				  & FS                 & 48.26\%          & 94.59\%          & 96.52\%          & 42.90\%                & 90.81\%                   & 96.48\%    \\
	\hline
	\multirow{5}{*}{$EAD_{{\ell}_{1}}$}
				  & \textbf{ContraNet} & \textbf{86.55\%} & \textbf{98.59\%} & \uit{94.54\%}    & \textbf{74.09\%}       & {\textbf{96.44\%}}        & \textbf{93.1\%} \\
				  & Trapdoor           & 45.36\%          & 1.22\%           & 83.51\%          & 42.54\%                & 0.36\%                    & 80.37\%     \\
				  & RR                 & 11.92\%          & 16.32\%          & \textbf{99.87\%} & 7.74\%                 & 8.24\%                    & \textbf{99.67\%}\\
				  & MagNet             & \uit{75.41\%}    & \uit{98.14\%}    & 85.58\%          & {\uit{73.22\%}}        & \uit{96.11\%}             & 83.49\%       \\
				  & FS                 & 49.75\%          & 64.57\%          & 5.05\%           & \uit{73.22\%}          & 35.16\%                   & 3.23\%   \\
	\hline
	\multirow{5}{*}{$C\&W_{{\ell}_{\infty}}$}
				  & \textbf{ContraNet} & \textbf{89.49\%} & \textbf{99.04\%} & \uit{99.75\%}    & {\uit{78.35\%}}        & \textbf{97.59\%}          & \textbf{99.75\%}  \\
				  & Trapdoor           & 25.39\%          & 4.94\%           & 82.73\%          & 22.86\%                & 4.2\%                     & 79.72\%       \\
				  & RR                 & 11.58\%          & 15.27\%          & \textbf{99.87\%} & 7.19\%                 & 7.43\%                    & \uit 99.71\% \\
				  & MagNet                 & \uit{82.82\%}    & \uit{96.39\%}    & 95.74\%      & \textbf{81.29\%}       & {\uit{93.36\%}}           & 94.45\%       \\
				  & FS                 & 50.79\%          & 41.57\%          & 17.51\%          & 22.85\%                  & 30.86\%                 & 13.25\%      \\			 
	\hlinew{1pt}
	\end{tabular}
	\begin{tablenotes}
		\footnotesize
		\item[*] For PGD, BIM, and C\&W, ${\ell}_{\infty}=8/255$, and for EAD, ${\ell}_{1}=8/255$, the same below.
		\item[$\diamond$] The \textbf{bolded} values are the best performance, and the \uit{underlined italicized} values are the second-best performance, the same below. 
	\end{tablenotes}}
\end{threeparttable}
\end{table*}

\begin{table*}[ht]
	\caption{TPR@FPR~5\% or 3\% of White-box Attack}
	\label{tab:TPR}
	\centering
	\setlength{\tabcolsep}{5mm}{
	\begin{tabular}{llcccccc}
	\hlinew{1pt}
	\multicolumn{2}{c}{Dataset}   & \cifar         & \multicolumn{1}{c}{\gtsrb} & \multicolumn{1}{c}{\mnist} & \cifar         & \multicolumn{1}{c}{\gtsrb} & \multicolumn{1}{c}{\mnist}\\ \hline
	Attack Method & Defense            & \multicolumn{3}{c}{\textbf{TPR@FPR5\%} $\uparrow$  }  & \multicolumn{3}{c}{\textbf{TPR@FPR3\%} $\uparrow$  }              \\ \hline
					
	\multirow{5}{*}{$PGD_{{\ell}_{\infty}}$}  
		 &\textbf{ContraNet}  & \textbf{92.6\%}  & \textbf{97.81\%} & \textbf{100\%}  & \textbf{83.55\%}       & \textbf{95.38\%}          & \textbf{100\%}         \\
		 & Trapdoor           & \uit{81.07\%}    & \uit{95.00\%}    & 99.87\%   & 59.32\%                & {\uit{94.8\%}}            & 99.87\% \\
		 & RR                 & 16.32\%          & 21.97\%          & \textbf{100\%}  & 59.32\%                & {\uit{94.8\%}}            & 99.87\%\\
		 & MagNet             & 76.68\%          & 44.26\%          & \textbf{100\%} & {\uit{75.19\%}}        & 39.77\%                   & 39.51\% \\
		 & FS                 & 65.50\%          & 90.76\%          & 14.81\%   & 57.67\%                & 84.82\%                   & 11.11\%\\
	\hline
	\multirow{5}{*}{$BIM_{{\ell}_{\infty}}$}  
	     & \textbf{ContraNet} & \textbf{93.62\%} & \uit{97.79\%}    & \textbf{100\%}   & \textbf{83.49\%}       & {\uit{94.70\%}}           & \textbf{100\%}         \\
		 & Trapdoor           & \uit{86.86\%}    & \textbf{97.98\%} & 99.50\%   & {\uit{75.28\%}}        & \textbf{97.44\%}          & 99.46\% \\
		 & RR                 & 13.92\%          & 15.31\%          & \textbf{100\%}  & 8.51\%                 & 11.05\%                   & 97.73\%   \\
		 & MagNet                 & 63.82\%          & 43.00\%          & \textbf{100\%}  & 62.42\%                & 39.56\%                   & 94.41\%   \\
		 & FS                 & 50.38\%          & 94.39\%          & 13.79\%  & 44.91\%                & 90.46\%                   & 12.07\% \\
	\hline
	\multirow{5}{*}{$EAD_{{\ell}_{1}}$}
	     & \textbf{ContraNet} & \textbf{89.12}\% & \textbf{99.07\%} & \uit{94.62\%}   & \textbf{76.96\%}       & {\uit{96.88\%}}           & 93.78\%          \\
		 & Trapdoor           & 51.24\%          & 1.23\%           & 84.17\%   & 48.12\%                & 0.39\%                    & 81.06\%   \\
		 & RR                 & 10.73\%          & 14.49\%          & \textbf{100\%}  & 7.11\%                 & 6.74\%                    & \textbf{99.96\%}   \\
		 & MagNet             & \uit{78.59\%}    & \uit{98.61\%}    & 86.85\%  & {\uit{76.65\%}}        & \textbf{97.21\%}          & 96.12\%  \\
		 & FS                 & 31.29\%          & 65.43\%          & 5.08\%  & {\uit{76.65\%}}        & 56.10\%                   & 3.28\%  \\
	\hline
	\multirow{5}{*}{$C\&W_{{\ell}_{\infty}}$}
		 & \textbf{ContraNet} & \textbf{92.51\%}        & \textbf{99.11\%}        & \textbf{99.92\%}  & {\uit{81.50\%}}        & \textbf{98.50\%}          & 99.92\%       \\
		 & Trapdoor           & 25.39\%                 &  5.06\%                 & 83.38\%   & 25.06\%                & 4.3\%                     & 80.4\%      \\
		 & RR                 & 11.58\%                 &  13.15\%                & \uit{99.72\%}  & 6.45\%                 & 5.85\%                    & \textbf{100\%} \\
		 & MagNet             & \uit{82.82\%}           &  \uit{96.82\%}          & 96.93\%   & \textbf{85.24\%}       & {\uit{94.41\%}}    & 73.75\%        \\
		 & FS                 & 50.79\%                 &  41.97\%                & 17.73\%   & 23.57\%                  & 31.13\%                   & 13.41\%        \\				
	\hlinew{1pt}
	\end{tabular}}
\vspace{-8pt}
\end{table*}

\noindent
\textbf{Detection performance.} Tab.~\ref{tab:TPR} reports the TPR@FPR5\% or 3\% to present \contranet's AE detection ability.
It is confirmed that \contranet can generalize well to different attacks for the same reason outlined in $Acc_{rob}$.
In addition, we can observe that ContraNet outperforms or is at least on par with other detection-based defenses. 
Especially when encountering \cifar dataset whose distribution is relatively complex, \contranet leads all other methods by a large margin.
For more results under ${\ell}_{2}$-norm, please refer to  Appendix~\ref{appendix:l2_detector_acc}.

\begin{figure}[t]
    \centering
    \includegraphics[width=0.8\linewidth]{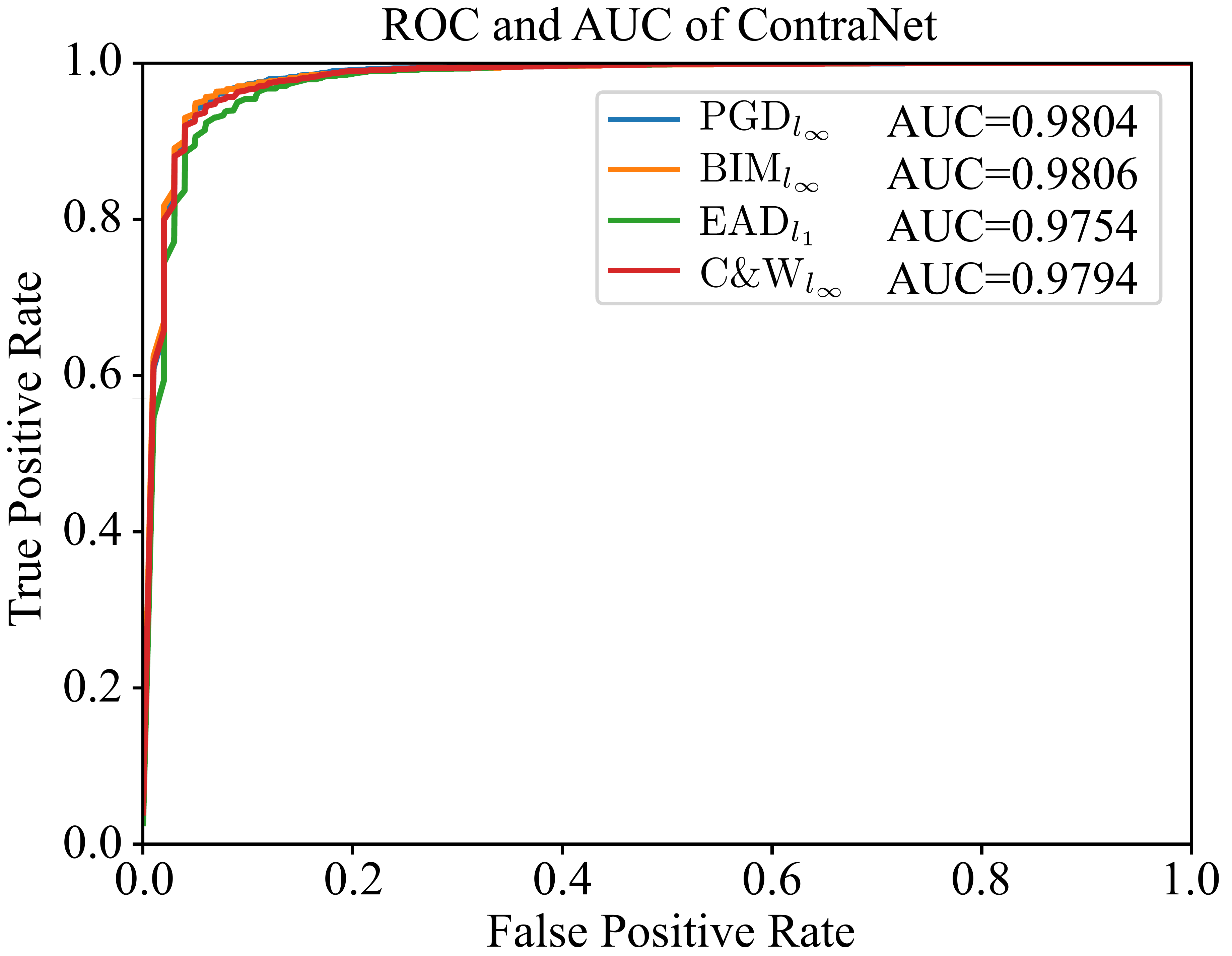}
    \caption{ContraNet's ROC and AUC. ROCs are generated under four attacks PGD$_{{\ell}_{\infty}}$, BIM$_{{\ell}_{\infty}}$, EAD$_{{\ell}_{1}}$, C\&W$_{{\ell}_{\infty}}$, along with the corresponding AUC (higher is better) demonstrated in the legend block.
    }
    \label{fig:roc}
	\vspace{-15pt}
\end{figure}

\noindent
\textbf{ROC curve and AUC.}
In Fig.~\ref{fig:roc}, we plot ROC curves together with AUC values for \cifar. TPR rises rapidly, with FPR increasing from low values.
This trendency indicates ContraNet can detect AE with high accuracy while keeping a low impact on legitimate samples.
Furthermore, AUC can reflect the overall performance with different thresholds.
ContraNet's AUCs have all been close to $1$, indicating it can detect AEs very well under different attacks.
ROC under $\ell_{2}$-norm is in Appendix~\ref{appendix:roc}.

\begin{figure}
	\centering
	\subfigure[False positives of ContraNet. First row: clean images misjudged as AEs. Second row: the corresponding synthetic images.]{
		\begin{minipage}{\linewidth} 
            \includegraphics[width=\textwidth]{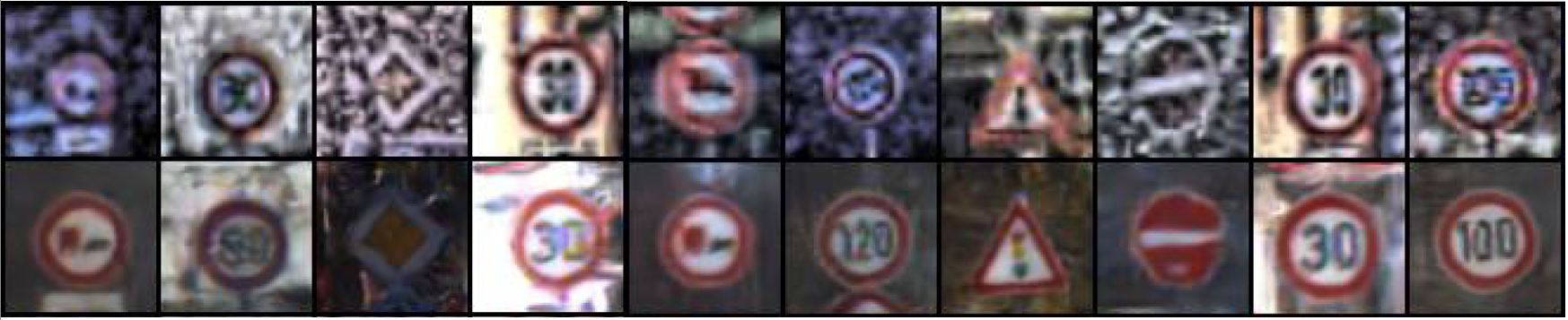} \\
		\end{minipage}
	}
	\subfigure[False negatives of ContraNet. First row: AEs generated by C\&W attack but misjudged as clean images by ContraNet. Second row: the corresponding synthetic images.]{
		\begin{minipage}{\linewidth}
			\includegraphics[width=\textwidth]{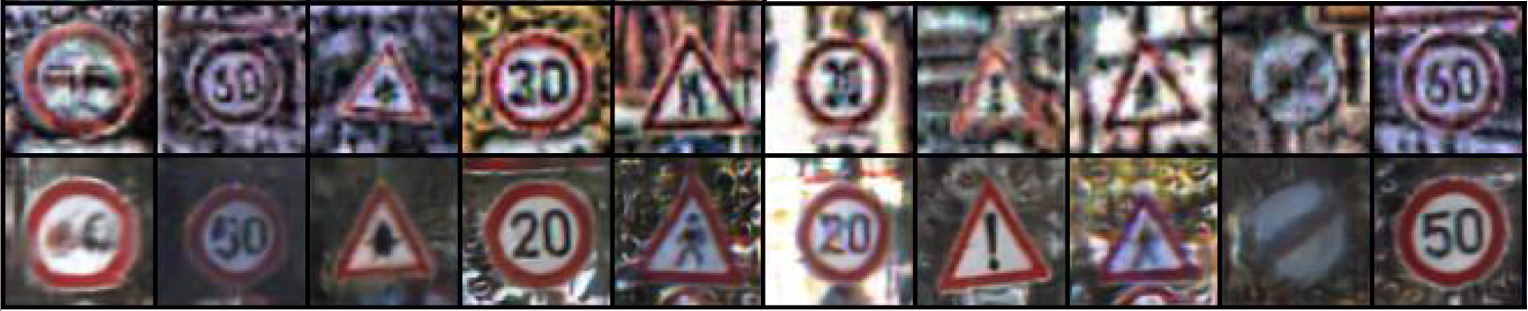} \\
		\end{minipage}
	}
	\caption{False positives and false negatives of ContraNet.}
	\label{fig:failurecase}
	\vspace{-8pt}
\end{figure}

\noindent
\textbf{Failure Cases.}
In Fig.~\ref{fig:failurecase}, we visualize representative failure cases (false positives and false negatives, respectively) of ContraNet. As can be observed, those legitimate input images that are misjudged as AEs (Fig.~\ref{fig:failurecase}~(a)) are corrupted to some degree, which is difficult to perceive even for humans. Interestingly, many of the AEs misjudged as clean inputs (Fig.~\ref{fig:failurecase}~(b)) have somehow lost their adversarial properties to humans as the semantics of the AEs are indeed close to that of the adversarial label. The above phenomena further demonstrate the effectiveness of ContraNet.

\subsection{ContraNet against AutoAttack}
\label{sec:ContraNet against AA attacks}

\emph{AutoAttack}~\cite{croce2020reliable} is designed for evaluating Adversarially Trained Classifier (ATC).
It is an ensemble of various PGD attacks and covers a wide range of attack settings (e.g., targeted/untargeted attacks and white-box/black-box attacks). 
As a benchmark, AutoAttack dedicates to providing a sufficient and impartial evaluation of adversarial defenses with auto-tuned hyperparameters. 
However, AutoAttack can only attack classifiers, but not detection-based methods.
Therefore, we follow the white-box setting demonstrated in Sec.~\ref{sec::exp_settings}, and use AEs generated by AutoAttack on each ATC from \cite{croce2020robustbench} for evaluation.
As shown in Tab.~\ref{tab:autoattack}, the addition of ContraNet increases the robustness of ATC by a significant margin on both clean samples and AEs.
To be more specific, compared with the vanilla classifier ($>95\%$ on \cifar), the standalone ATC induces a poor $Acc$.
When equipped with \contranet, the accuracy increases to $>91.4\%$ indicated by $Acc_{dec}$.
This is reasonable because \contranet can distinguish misclassification clean samples caused by ATC, thus increasing accuracy on clean samples.
When it comes to AEs, \contranet can filter out AEs that have fooled ATC, resulting in a 15.87\% uplift on $Acc_{rob}$, as shown in Tab.~\ref{tab:autoattack}'s the 3rd and 4th columns.

\begin{table}[t]
	\caption{ATC with ContraNet against AutoAttack on \cifar}
	\label{tab:autoattack}
	\centering
	\begin{tabular}{lcccc}
	\toprule
	\multirow{2.5}{*}{ATC Methods} & \multicolumn{1}{c|}{$Acc$} & \multicolumn{1}{c|}{$Acc_{dec}$} & \multicolumn{2}{c}{\makecell[c]{$Acc_{rob}$ \\ (under AutoAttack)}} \\
	\cmidrule{2-5} 
	 \multicolumn{1}{c}{}  &
	  \multicolumn{1}{c|}{ATC} &
	  \multicolumn{1}{c|}{\makecell[c]{ATC\\+ ContraNet}} &
	  \multicolumn{1}{c}{ATC} &
	  \multicolumn{1}{c}{\makecell[c]{ATC\\+ ContraNet}} \\ \midrule
	G.2020Uncovering~\cite{gowal2020uncovering}                     & 85.28\% & \multicolumn{1}{|c|}{91.73\%} & 57.14\%        & 84.96\%       \\
	R.2021Fixing\_28~\cite{rebuffi2021fixing}                   & 87.32\% & \multicolumn{1}{|c|}{91.43\%} & 57.42\%        & 82.54\%       \\
	R.2021Fixing\_70~\cite{rebuffi2021fixing}                   & 88.97\% & \multicolumn{1}{|c|}{91.40\%} & 57.33\%        & 83.58\%       \\
	S.2020Hydra~\cite{Sehwag2020Hydra}                         & 88.53\% & \multicolumn{1}{|c|}{91.69\%} & 64.46\%        & 80.33\%       \\ \bottomrule	\end{tabular}
\end{table}

\begin{figure}[t]
    \centering
    \includegraphics[width=0.9\linewidth]{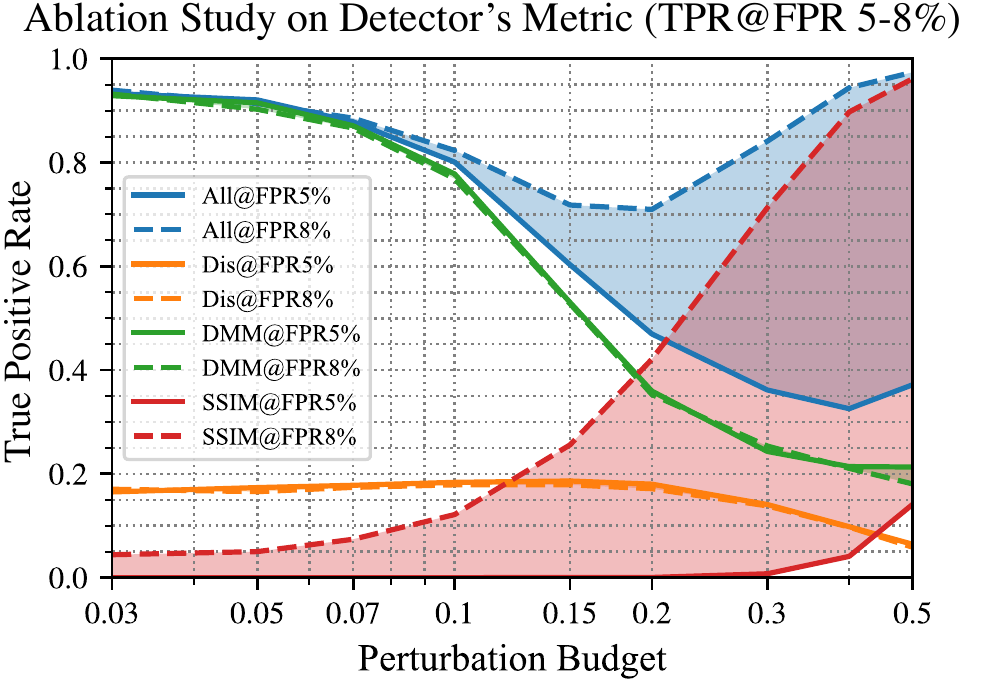}
    \caption{Ablation study for Similarity measurement model on TPR@FPR5\% to 8\%. AEs are generated by PGD$_{{\ell}_{\infty}}$ on \cifar.}
    \label{fig:ablation_study_fpr5}
	\vspace{-15pt}
\end{figure}	

\subsection{Ablation Study}
\label{sec::ablation_study}
In this section, we perform an ablation study on each component of the Similarity measurement model, i.e., \dmm,  \dis and $\ssim$, and analysis their detection ability under a wide scope of perturbation budgets.
We show the \textit{TPR@FPR 5-8\% v.s. Perturbation budgets} curves keeping the thresholds as in Sec.~\ref{sec:ContraNet against white-box attacks} (Fig.~\ref{fig:ablation_study_fpr5}).

\noindent
\textbf{Deep Metric Model} (\dmm) is the most effective detector when the adversarial perturbation falls in small scope ($< 10^{-1}$, green lines in Fig.~\ref{fig:ablation_study_fpr5}). 
Especially when the adversarial perturbation ranges from 0.03 to 0.07, \dmm alone can achieve an over 90\% TPR.
The detection ability of the Similarity measurement model relies on \dmm for FPR@5\%-8\% situations in this perturbation range.
However, when the perturbation grows, \dmm's performance degrades sharply.
It is not hard to reason about this observation.
When the adversarial perturbation is larger than 0.1 on ${\ell}_{\infty}$-norm
, obvious image quality degrade occurs.
The noise-like (non-semantic) perturbation will result in the loss of the image's original semantic information.
Since the \dmm has no access to such distorted images during training, it performs poorly.   

\noindent {\boldmath{{$\ssim$}}} is responsible for detecting AEs with extra-large adversarial perturbations.
$\ssim$ is not a learning-based similarity metric
and its judgment criterion can be consistent with human perception~\cite{sara2019image}.
Therefore, $\ssim$ can hardly be broken with the same perturbation budgets as DNN-based metrics.
We find almost no effect of $\ssim$ on the clean sample's accuracy when we set the FPR to 5\%. However, this also limits $\ssim$'s ability to detect AE (see Fig.~\ref{fig:ablation_study_fpr5}).
$\ssim$'s detection ability can show up when relaxing the FPR a bit.
If we relax FPR from 5\% to 8\%, the Detector's Acc obtained by $\ssim$ will increase to 40\% for $\epsilon=0.2$, and this number will rise to 80\% for $\epsilon=0.5$, as shown by red shadow lines.
Such increases contribute much to the overall Detector's Acc, as indicated by the blue shadow.

\noindent\textbf{Discriminator.}
We visualize \dis's TPR curves by orange lines in Fig.~\ref{fig:ablation_study_fpr5}.
\dis assists \dmm against middle-level adversarial perturbation (0.1-0.2 under ${\ell}_{\infty}$-norm).
The adding of \dis acts as a relay linking \dmm with $\ssim$.
The overall performance is better than using any component alone, as shown by blue lines in Fig.~\ref{fig:ablation_study_fpr5}.
Although \dis has a limited effect on the white-box attacks, \textit{the real power of 
\dis is its robustness against adaptive attacks}, as analyzed in Sec.~\ref{sec:Adaptive}.

In conclusion,  \dmm, \dis, and $\ssim$ detect AE using different mechanisms; thus, their cooperation contributes to the overall performance of \contranet.
Note that each detector's performance can be configured relying on the actual situation.

\subsection{Comparison with Other Detection Methods}

\noindent
\textbf{Trapdoor.}
The detection principle of Trapdoor is to induce the adversary to generate AEs fall into pre-designed trap door(s) by retraining the classifier with stamped samples. We observe that this strategy works well with gradient-based attacks, such as PGD and BIM, where the perturbation is relatively large. 
In terms of optimization-based attacks like C\&W, and EAD,  the small perturbation constraint remains in their objective functions.
These attacks tend to find optimal alternatives other than the trap door to fool the classifier.

\noindent
\textbf{RR.} RR is an adversarial training detection framework, which gains better performance for seen attacks, i.e., PGD  and similar BIM attacks but this advantage will vanish when encountering attacks not included in the training process, like C\&W or EAD.

\noindent
\textbf{MagNet and FS.}  As two early detection-based defenses, MagNet and FS display inconsistent detection ability across  datasets and attacks. For example, MagNet is vulnerable to BIM attacks, and FS performs poorly on \cifar.

%% file: 7.adaptive_attack_evaluation.tex
\section{Adaptive Attacks}
\label{sec:Adaptive}

In this section, we switch our role from a defender to an attacker. To best utilize ContraNet's knowledge, we first give a detailed adaptive objective loss function design; then, we try to break ContraNet using three types of adaptive attacks.
The first two adaptive attacks are based on public-known strong iterative attacks, PGD~\cite{madry-PGD} and C\&W~\cite{carlini2018towards}. To further evaluate ContraNet's robustness, we employ a rising adaptive attack benchmark, \textit{Orthogonal PGD}~\cite{bryniarski2021evading}, which focuses on breaking detection-based defenses.

\subsection{Experimental Settings}
\label{sec::adaptive_exp_set ting}
\noindent\textbf{Adaptive attack.} 
In the adaptive attack setting, an adversary has \textit{complete knowledge} of the classifier and the defense scheme. Therefore, the adversary can develop an \textit{adaptive attack} to fool both simultaneously.

\noindent\textbf{Dataset.}
All experiments are conducted on \cifar, which serves as a standard task by several public robustness test-benches~\cite{robustml,croce2020robustbench}.

\noindent\textbf{Evaluation Metrics.}
The evaluation metrics are consistent with those introduced in Sec.~\ref{sec::exp_settings}.
Additionally, we demonstrate the $Acc_{rob}$  \textit{ v.s. Perturbation budget} curve for \contranet in order to further evaluate its effectiveness against various adversarial capabilities. The perturbation budget represents the upper limit of the adversarial capabilities.

\subsection{Customizable Adaptive Objective Loss Function}
\label{sec::adaptive_objective_losses}
As evidenced in \cite{CarliniEvaluatingAdversarialRobustness2019,carlini2017adversarial}, it is critical to choose the proper loss function for adaptive attacks.
On the one hand, an adaptive attack aims to  simultaneously bypass the detector and fool the classifier.
On the other hand, an overly complex objective loss function will complicate the optimization process resulting in failure attacks or suboptimal solutions.
Notice that only the classification loss can mislead the prediction result. 
Therefore, the attacker should always consider the classification loss.  The attackers may not consider all the losses items in the defense. Instead, they can focus on the defense's weakest point. 

\begin{figure}[t]
    \centering
    \includegraphics[width=\linewidth]{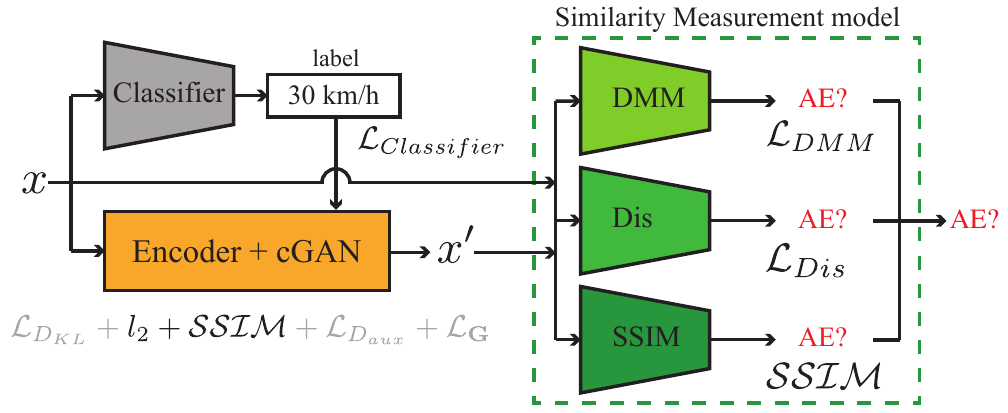}
    \caption{Loss terms in ContraNet that can be attacked adaptively. We drop $\mathcal{L}_{D_{KL}}$ for its little impact on adaptive attack. We drop the $\mathcal{L}_{\mathbf{G}}$ and $\mathcal{L}_{D_{aux}}$ because $D_{\Phi}$ and $D_{aux}$ are only used during training.}
    \label{fig:adaptive_loss_components}
	\vspace{-15pt}
\end{figure}

We re-depict \contranet's components with their training losses in Fig.~\ref{fig:adaptive_loss_components}.
ContraNet's loss functions can be further fine-grained into \textit{Deep Metric Model's loss} ($\mathcal{L}_{DMM}$, use Cross Entropy here),  \textit{Discriminator's loss} ($\mathcal{L}_{Dis}=ReLU(1-Dis(x))$), $\mathcal{SSIM}$~\cite{sara2019image} and ${\ell}_{2}$.
Adaptive attacks are conducted by varying the above loss items from targeting the single one to combinations, as shown in Eq.~\eqref{eq:loss_all} --~\eqref{eq:SSIM}.
Among these objectives, Eq.~\eqref{eq:loss_all} covers all losses used to train  ContraNet;
Eq.~\eqref{eq:loss_ssim_dis_dml} covers all three functions used in the Similarity measurement model during inference.
In addition, other objectives (e.g., Eq.~\eqref{eq:loss_dml} --~\eqref{eq:SSIM}) are also promising due to their simpler optimization process.

\begin{small}
\vspace{-15pt}
\begin{align}
	\mathcal{L}_{ContraNet_{1}} &= {\ell}_{2} + \mathcal{SSIM} + \mathcal{L}_{DMM} +\mathcal{L}_{Dis}\label{eq:loss_all}\\
	\mathcal{L}_{ContraNet_{2}} &= \mathcal{SSIM} + \mathcal{L}_{DMM} + \mathcal{L}_{Dis}\label{eq:loss_ssim_dis_dml} \\
	\mathcal{L}_{ContraNet_{3}} &= \mathcal{L}_{DMM} + \mathcal{L}_{Dis} \label{eq:loss_dml_dis}\\
	\mathcal{L}_{ContraNet_{4}} &= \mathcal{SSIM} + \mathcal{L}_{DMM} \label{eq:loss_ssim_dml}\\
	\mathcal{L}_{ContraNet_{5}} &= \mathcal{SSIM} + \mathcal{L}_{Dis} \label{eq:loss_ssim_dis}\\
	\mathcal{L}_{ContraNet_{6}} &= \mathcal{L}_{DMM}\label{eq:loss_dml}\\
	\mathcal{L}_{ContraNet_{7}} &= \mathcal{L}_{Dis}\label{eq:loss_dis}\\
	\mathcal{L}_{ContraNet_{8}} &= \mathcal{SSIM}\label{eq:SSIM}
\end{align}
\vspace{-15pt}
\end{small}

\noindent\textbf{Objective loss function for PGD adaptive attack.}
To conduct adaptive attack based on PGD attack, we modify the objective loss function as following:
\begin{align}\label{eq:adaptive_pgd}
	\mathcal{L}_{PGD}^\prime &= \mathcal{L}_{Classifier} +\underbrace{\lambda \cdot \mathcal{L}_{ContraNet_{i}}}_{where\ i = {1, ..., 7}}\text{,}
\end{align}
where the $\mathcal{L}_{Classifier}$ is the same as original {\it{PGD}} to attack the classifier, and $\lambda \cdot \mathcal{L}_{ContraNet_{i}}$ will try to evade ContraNet.
Notably, Orthogonal-PGD is a variant of PGD that has the same adaptive objective loss function as PGD.

\noindent
\textbf{Objective loss function for C\&W adaptive attack.}
As for C\&W attack, 
we keep the original C\&W objective loss items proposed in\cite{carlini2018towards}, while introducing the $\mathcal{L}_{ContraNet_{i}}$ 
to generate AEs to evade  ContraNet.
As stated in Eq.~\eqref{eq:adaptive_cw}, the first item of $\mathcal{L}_{C\&W}$ minimizes the distance between $x$ and $x_{adv}$.
The second item, $c\cdot f(x_{adv}),$ aims at deceiving the classifier.
Finally, $\lambda \cdot \mathcal{L}_{ContraNet_{i}}$ is to evade the ContraNet. 
\begin{multline}
	\label{eq:adaptive_cw}
	\mathcal{L}_{C\&W}^\prime =
	 \underbrace{\overbrace{\left \| x_{adv} - x \right \|^{2}}^{min\ perturbation} + \overbrace{c\cdot f(x_{adv})}^{ misclassificated}}_{\mathcal{L}_{C\&W}} \\
	 + \underbrace{\lambda \cdot \mathcal{L}_{ContraNet_{i}}}_{where\ i={1,...,7}}
\end{multline}

\subsection{Performance against PGD Adaptive Attacks}
\label{sec::pgd_adaptive}
\noindent
\textbf{PGD experimental settings.}
We perform the PGD targeted adaptive attacks on various objective loss functions defined in Sec.~\ref{sec::adaptive_objective_losses}.
We varies the adversary perturbation budget from 0.01 to 0.5 with ${\ell}_{\infty}$-norm, to cover a large range of adversarial capabilities.
The iteration steps are set to 200 (we also tested with 400 to verify the attacker's ability, as recommended in \cite{CarliniEvaluatingAdversarialRobustness2019}).
We set $\lambda=1$ in Eq.~\eqref{eq:adaptive_pgd}.

\begin{figure}[t]
    \centering
    \includegraphics[width=0.9\linewidth]{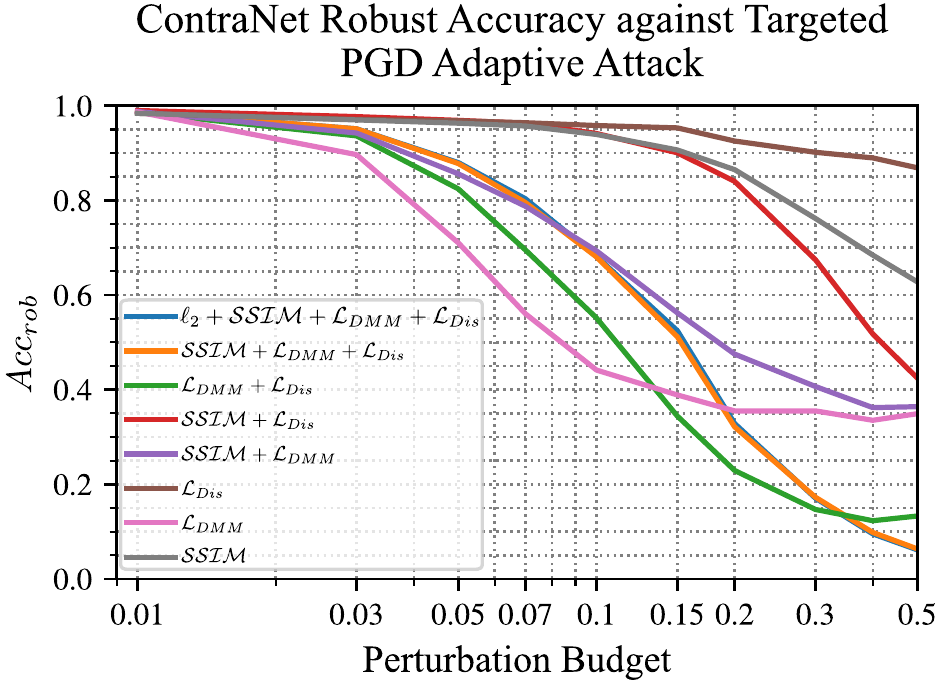}
    \caption{Adaptive PDG attacks on ContraNet according to multiple objectives under different perturbation budgets. 
	}
    \label{fig:adaptive_acc_curves_targeted}
\vspace{-10pt}
\end{figure}

\noindent
\textbf{Robust accuracy v.s. perturbation budget curves.} Fig.~\ref{fig:adaptive_acc_curves_targeted} summarizes the $Acc_{rob}$ varying with different perturbation budgets associated with various objective loss functions. 
We analyze Fig.~\ref{fig:adaptive_acc_curves_targeted} as follows:
1) When the perturbation is under 0.1, 
with $\mathcal{L}_{DMM}$, PGD can attack the whole model to the lowest $Acc_{rob}$.
This may be because that \dmm plays a significant role mostly in detecting the AEs with small perturbations, as stated in Section~\ref{sec::ablation_study}.
2) The line for $\mathcal{L}_{DMM}$ flattens out as the perturbation budget grows (> 0.1).
This might result from \dis and $\mathcal{SSIM}$ starting to work, and only attack \dmm can not degrade \dis and $\mathcal{SSIM}$'s performance. 
Therefore, to further degrade ContraNet's performance, we have to take \dis into consideration.
3) By comparing the green and pink lines, we can find that as the perturbation budget grows, attacking $\mathcal{L}_{DMM}+\mathcal{L}_{Dis}$ is more effective than $\mathcal{L}_{DMM}$. 
4) With the perturbation increasing continually, $\mathcal{SSIM}$'s detection capabilities begin to show.
In this case, merely attacking \dmm and \dis cannot influence $\mathcal{SSIM}$'s performance.
Consequently, (${\ell}_{2}+$) $\mathcal{SSIM}+\mathcal{L}_{DMM}+\mathcal{L}_{Dis}$ shown by the yellow (blue) line becomes the most effective attack. 
5) Other objective loss functions are either too simple or have not targeted the weakest point of ContraNet, therefore, less effective.

\begin{figure}[t]
	\centering
	\includegraphics[width=\linewidth]{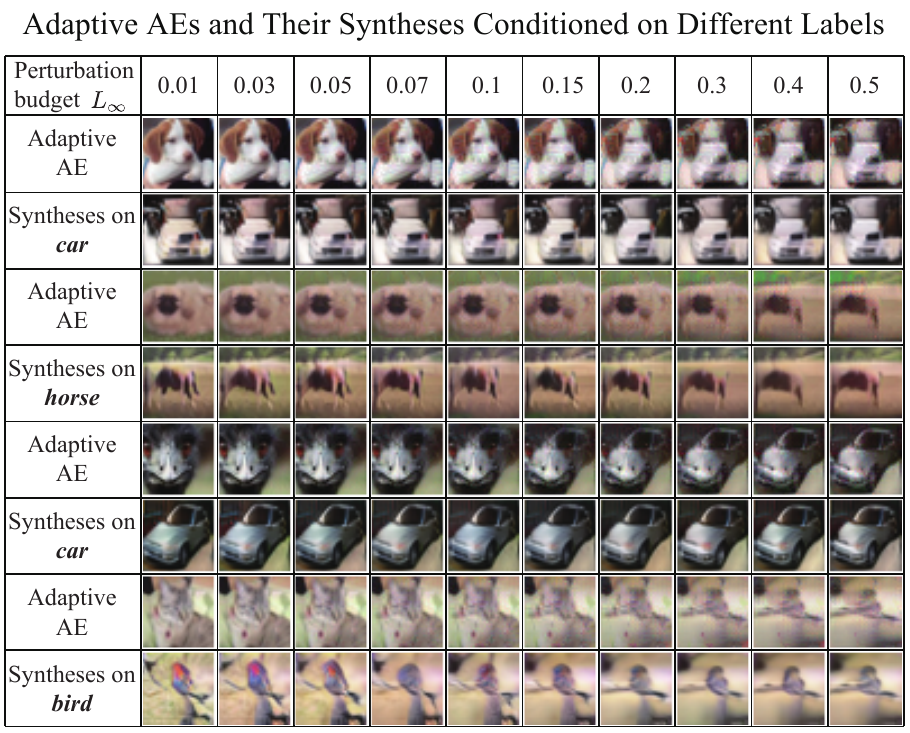}
	\caption{Adaptive AE attacks under different perturbation budgets. To bypass ContraNet, the generated AEs resemble the target class, thereby gradually losing malignity. We call it AE's ``self-defeat".
	}
	\vspace{-15pt}
	\label{fig:adaptive_exampels}
\end{figure}

\noindent
\textbf{AE's ``self-defeat''.}
According to Corollary~\ref{cor:feature}, AEs will lose their malignity if their semantic information changes.
We call this phenomenon AE's ``self-defeat''.
Therefore, it is beneficial to trigger AE's self-defeat, when defending adaptive attacks.

We notice that ContraNet's detection mechanism can result in adaptive AEs losing their malignancy.
As shown in Fig.~\ref{fig:adaptive_exampels}, all adaptive AEs will end up being similar to their synthetic images, and appear the semantic features of its targeted label.
Therefore, these AEs are ``self-defeated''.

The observed self-defeat phenomenon is related to \contranet's detection mechanism.
Consider a successful AE that fools the classifier to a targeted label and bypasses the ContraNet.
To evade ContraNet, 1) $\mathcal{SSIM}$ requires AE perceptually similar to the synthesis; 2) \dmm requires AE semantically similar to the synthesis; and 3) \dis requires AE with high visual quality.
Therefore, this successful AE should be very similar to its synthetic image.
Notice that the synthesis from ContraNet always keeps the visible semantic feature of the targeted class.
Hence, a successful AE will resemble a clean sample from that class.

\begin{figure}[t]
	\centering
	\includegraphics[width=0.85\linewidth]{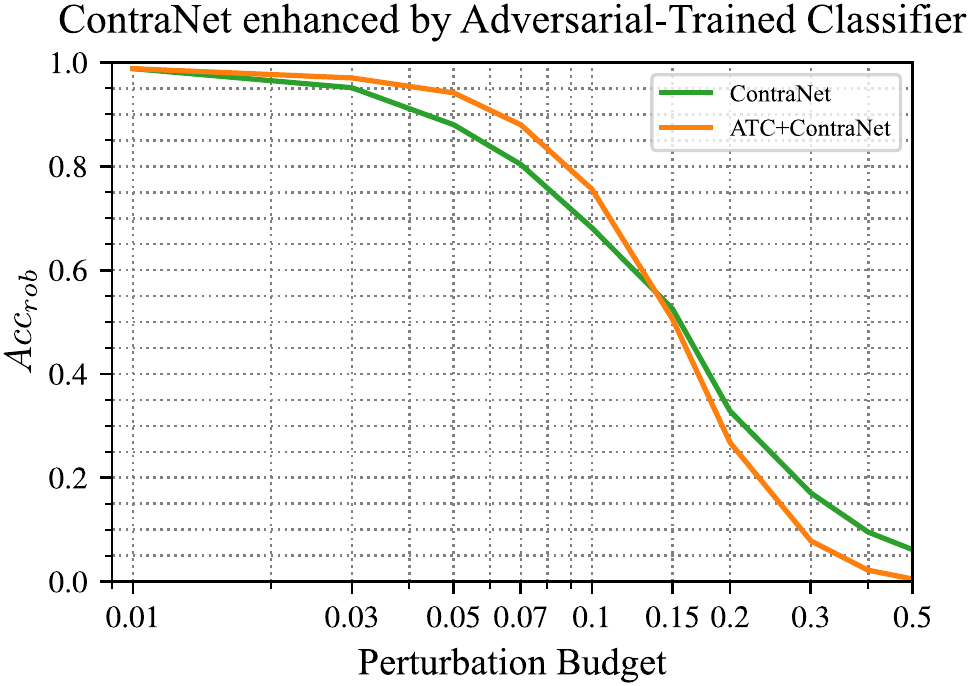}
	\caption{Robust accuracies under adaptive PGD attack on \cifar. \textit{ContraNet} indicates applying ContraNet on normal-trained classifier, while \textit{ATC+ContraNet} indicates combining ContraNet with Adversarially-Trained Classifier (Gowal2020Uncovering~\cite{gowal2020uncovering} from RobustBench~\cite{croce2020robustbench}).
	}
	\label{fig:atc_robust_acc_curves}
\vspace{-15pt}	
\end{figure}

\noindent
\textbf{Enhancing ContraNet's performance with adversarial training.} Incorporating with Adversarial-Trained Classifier (ATC) can further improve the robustness performance of ContraNet. 
We directly use an ATC, Gowal2020Uncovering~\cite{gowal2020uncovering}, from RobustBench~\cite{croce2020robustbench}, to incorporate with ContraNet forming a robust classifier (\textit{ATC + ContraNet}).  
Fig.~\ref{fig:atc_robust_acc_curves} shows this robust classifier's $Acc_{rob}$ v.s. \textit{perturbation budget} curve in orange.
The green line demonstrates the combination of the normal-trained classifier and ContraNet, denoted as \textit{ContraNet}. 
Clearly, when the perturbation is relatively small, \textit{ATC + ContraNet} outperforms the non-adversarial classifier. 
However, when the adversarial perturbation grows larger than 0.1, \textit{ContraNet} tends to overtake \textit{ATC + ContraNet}.

\begin{figure}[t]
	\centering
	\includegraphics[width=\linewidth]{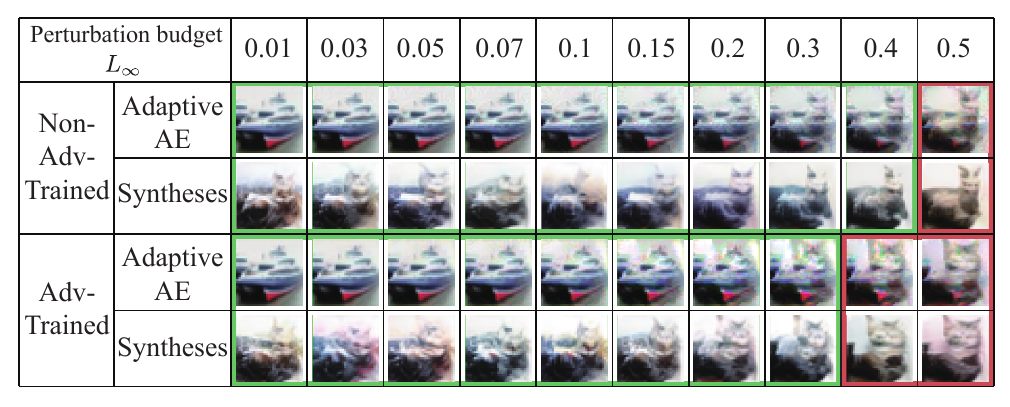}
	\caption{AEs and synthetic images on Adversarially (Adv) / Normal (Non-Adv) -Trained classifier + ContraNet against adaptive PGD attack on \cifar.
	The targeted class is ``cat''.
	Cases in green boxes are rejected by ContraNet.
    Cases in red boxes can bypass ContraNet.
    ${\ell}_{\infty}=0.4$ shows that although ATC+ContraNet performs poorly on rejection than normal classifier+ContraNet, the corresponding AE are more likely to be self-defeated.
	}
	\label{fig:adv-trained_adaptive_example}
	\vspace{-15pt}
\end{figure}

Actually, ATC helps ContraNet defend against large adversarial perturbation by triggering AE's self-defeat.
\cite{santurkar2019image, shafahi2019adversarial} have stated that when the perturbation goes larger, the gradients of ATC will become interpretable.
Such property of ATC can help trigger AE's self-defeat.
To further explain this phenomenon, we give concrete examples in Fig.~\ref{fig:adv-trained_adaptive_example}.
Despite the attacker requiring lower perturbation budgets, ${\ell}_{\infty} = 0.4$, to breach \textit{ATC + ContraNet} than a normal-trained classifier, the adaptive AE generated on \textit{ATC + ContraNet} is more like a cat than the one produced on \textit{normal-trained classifier + ContraNet}. 
As a result, adversarial training and ContraNet can benefit each other in triggering AE's self-defeat.

To sum up, for small perturbations, ATC enhances ContraNet's performance by increasing the $Acc_{rob}$ directly.
While for large perturbations, ATC supports ContraNet to lure the adaptive AE to be self-defeated, i.e., look like the clean sample from the targeted class.

\subsection{Performance against C\&W Adaptive Attacks}\label{sec::cw_adaptive}
\noindent
\textbf{C\&W experimental settings.} We employ the adaptive losses designed for \contranet in Sec.~\ref{sec::adaptive_objective_losses} with C\&W's objectives as the final adaptive objectives.
We set the iteration steps to 1000 (verified to converge the optimal attack), the initial constant to 0.01, and binary search steps to 9.
We employ Adam~\cite{Bengio2014Adam} with a 0.005 learning rate as the optimizer to search for the AEs.
As for the hyperparameter $\lambda$ in Eq.~\eqref{eq:adaptive_cw}, we try $\lambda=1$ and $\lambda=c$ in our experiments.~\footnote{$c$ in Eq.~\eqref{eq:adaptive_cw} is binary searched by C\&W}

\noindent
\textbf{Robustness accuracy against C\&W adaptive attacks.} Tab.~\ref{tab:cw_adaptive_targeted} demonstrates the performance of ContraNet against targeted and untargeted C\&W attacks.
As can be seen, under C\&W attacks, ContraNet can still keep a 55.28\% $Acc_{rob}$.
Moreover, different $\mathcal{L}_{ContraNet_{i}}$ 
tend to result in similar $Acc_{rob}$.
This may be due to $\mathcal{L}_{C\&W}$, which optimizes AE's perturbation to be small.
From Sec.~\ref{sec::ablation_study} and \ref{sec::pgd_adaptive}, we notice that \dmm is the most effective component to small perturbation.
Therefore, except $\mathcal{L}_{DMM}$, other loss items in the adaptive C\&W attacks only have little impact on $Acc_{rob}$.

\begin{table}[t]
\vspace{-10pt}
	\caption{Robust Accuracy of ContraNet against Targeted C\&W Adaptive Attack}
	\label{tab:cw_adaptive_targeted}
	\centering
	\begin{threeparttable}
	\begin{tabular}{lc}
	\hline
	\textbf{Objective Loss Function} & $Acc_{rob}$ \\ \hline
	$\mathcal{L}_{C\&W}+\mathcal{L}_{DMM}$                & 56.30\%                                   \\
	$\mathcal{L}_{C\&W}+\mathcal{L}_{Dis}$                 & 56.37\%                                   \\
	$\mathcal{L}_{C\&W}+\mathcal{L}_{Dis}+\mathcal{L}_{DMM}$                                & 56.25\%                                   \\
	$\mathcal{L}_{C\&W}+\mathcal{L}_{Dis}+\mathcal{L}_{DMM}+\mathcal{SSIM}$                     & 56.09\%                                   \\
	$\mathcal{L}_{C\&W}+l_{2}+\mathcal{L}_{Dis}+\mathcal{L}_{DMM}+\mathcal{SSIM}$                                 & \textbf{55.28\%}                                   \\
	$\mathcal{L}_{C\&W}+\lambda\cdot (l_{2}+\mathcal{L}_{Dis}+\mathcal{L}_{DMM}+\mathcal{SSIM})$                     & 56.37\%                                   \\
	\hline
	\end{tabular}
	\begin{tablenotes}
		\footnotesize
		\item[$\diamond$] The \textbf{bolded} value indicates the worst performance. 
	\end{tablenotes}
	\end{threeparttable}
\vspace{-10pt}
\end{table}

\subsection{Performance against Orthogonal-PGD Adaptive Attacks}
\noindent
\textbf{Orthogonal-PGD experimental settings.} Orthogonal Projected Gradient Descent (Orthogonal-PGD)~\cite{bryniarski2021evading} is the most recent proposed benchmark for AE detection defenses.
There are two attack strategies in Orthogonal-PGD, \textit{Selective strategy } (Select) and \textit{Orthogonal strategy} (Orth).
In the \textit{Selective strategy}, Orthogonal-PGD update the input by selectively use the perturbation generated by the classifier or that of the detector to avoid the over-optimization on either of the two.
The \textit{Orthogonal strategy} only keeps the orthogonal component of the gradient from both the classifier and the detector when optimizing to prevent them from disturbing each other.
We adopt these two strategies to attack ContraNet from three aspects:
\begin{itemize}[leftmargin=*, itemsep=3pt]
	\item \textit{Adopting Orthogonal-PGD to ContraNet directly.} 
	We perform the two optimization strategies of Orthogonal-PGD, selective and orthogonal on ContraNet, respectively and report the results.
	\item \textit{Combining Orthogonal-PGD with adaptive loss designs.}
	Sec.~\ref{sec::pgd_adaptive} shows that attacking \dmm is the most effective when the perturbation is relatively small.
	Since we use Orthogonal-PGD with ${\ell}_{\infty}=0.01$ and $8/255$, we strengthen Orthogonal-PGD by letting it attack \dmm alone.
	\item \textit{Explore how adversarial training helps ContraNet}.
	We also test ContraNet's performance by letting it work with Adversarial-Trained Classifier (ATC) (as in Sec.~\ref{sec::pgd_adaptive}) to further explore how adversarial training techniques help ContraNet.
\end{itemize}
We set the optimization step to 1000 for all the experiments.
We consider two adversarial perturbation budgets, ${\ell}_{\infty}=0.01$ and ${\ell}_{\infty}=8/255$, and report the $Acc_{rob}$~@FPR5\% and @FPR50\% as the evaluation metrics (following Orthogonal-PGD, performance @FPR50\% serves as the worst case).\footnote{In~\cite{bryniarski2021evading}, the authors report the successful rate, which is equivalent to ours $Acc_{rob}$, and the relationship of these two metrics is depicted in Eq.~\eqref{eq:robust_acc}}

\noindent
\textbf{Robust Accuracy against Orthogonal-PGD.}
As shown in Tab.~\ref{tab:opgd}, ContraNet outperforms four baselines by a considerable margin cross both ``Select'' and ``Orth'' attacks for all attack scenarios (ContraNet, ATC + ContraNet).
For the worst-case test, ContraNet can still keep 38.1\% (Orth) $Acc_{rob}$ when ${\ell}_{\infty}=8/255$.
Further, incorporating the adversarial training technique can significantly improve ContraNet's performance from 38.1\% (Orth) to 89.7\% (Select).
The detection mechanism guarantees the ContraNet's robustness against Orthogonal-PGD:
since the synthesis's semantic information is highly dependent on the classifier's discriminative features, one can hardly alter
this dependency
with bounded perturbation upon input image.

\subsection{Summary of the Adaptive Attacks}
From the attacker's point of view, Orthogonal-PGD achieved a higher attack success rate than the prior conducted PGD and C\&W adaptive attacks (Sec.~\ref{sec::pgd_adaptive} and \ref{sec::cw_adaptive}).
However, the attack performance of Orthogonal (Orth) and Selective (Select) are close to each other.
To analyze this phenomenon, we check the gradients for both classifier and detector, and find out that gradients of classifier and ContraNet is almost orthogonal to each other.
In this case, either Orthogonal or Selective will optimize the AE similarly.
Therefore, there is only a slight difference between the two results.
The nearly orthogonal gradients provide an explanation for the difficulty of breaching both the \contranet and the classifier simultaneously.

From the adaptive objective loss function's point of view, focusing on the weakest point of \contranet can simplify the optimization process, and get a better attack success rate.
By comparing the performance of \contranet on
$\mathcal{L}_{DMM}$ and \contranet on $\mathcal{L}_{ContraNet_{1}}$ in Tab.~\ref{tab:opgd}, it is clear that $\mathcal{L}_{DMM}$ as the objective function will further decline ContraNet's $Acc_{rob}$ than take all loss items, $\mathcal{L}_{ContraNet_{1}}$, into consideration.
This verifies our intention to replace  $\mathcal{L}_{ContraNet_{1}}$ with $\mathcal{L}_{DMM}$.

The adding of ATC can enhance ContraNet's performance.
As shown in Tab.~\ref{tab:opgd}, the performance for ATC + ContraNet is the best for all test settings.
Therefore, it is promising to combine ContraNet with ATC to improve robustness.
We leave this as future work.
\begin{table}[t]
	\caption{Robust Accuracy under
	 Orthognal-PGD Attack}
	\label{tab:opgd}
	\centering
	\begin{threeparttable}
	\begin{tabular}{@{}llcccc@{}}
  	\hlinew{1pt}
	\multirow{3.5}{*}{\textbf{Attack}} &
	\multirow{3.5}{*}{\textbf{Defense}} &
	\multicolumn{2}{c}{$L_{\infty}=0.01$} &
	\multicolumn{2}{c}{$L_{\infty}=8/255$} \\ \cline{3-6} 
	&
	&
	\multicolumn{1}{l}{\begin{tabular}[c]{@{}l@{}}$Acc_{rob}$\\ @FPR5\%\end{tabular}} &
	\multicolumn{1}{l}{\begin{tabular}[c]{@{}l@{}}$Acc_{rob}$\\ @FPR50\%\end{tabular}} &
	\multicolumn{1}{l}{\begin{tabular}[c]{@{}l@{}}$Acc_{rob}$\\ @FPR5\%\end{tabular}} &
	\multicolumn{1}{l}{\begin{tabular}[c]{@{}l@{}}$Acc_{rob}$\\ @FPR50\%\end{tabular}} \\
	\hline
	\multirow{7}{*}{Orth}
		& \textbf{ContraNet}\tnote{*}  & \uit{85.4\%} & \uit{97.3\%} & \uit{63.7}\% & \uit{83.0\%} \\
		& \textbf{ContraNet}\tnote{$\dagger$}
			& 79.3\%     & 95.7\%          & 38.1\%          & 77.9\% \\
		& \textbf{ContraNet}\tnote{$\dagger$}~\textbf{+ATC}
			& \textbf{93.7\%} & \textbf{99.2\%} & \textbf{89.8}\% & \textbf{94.7\%} \\
		& Trapdoor\cite{Shan2020GottaCA} & 0\%           & 7.0\%             & 0\%             & 8.0\%             \\
		& DLA'20\cite{sperl2020dla} & 62.6\%        & 83.7\%          & 0\%             & 28.2\%          \\
		& SID'21\cite{tian2021detecting} & 6.9\%         & 23.4\%          & 0\%             & 1.6\%           \\
		& SPAM'19\cite{liu2019detection}     & 1.2\%         & 46.0\%            & 0\%             & 38.0\%            \\
	\hline
	\multirow{6}{*}{Select}
		& \textbf{ContraNet}\tnote{*} & \uit{85.4\%} & \uit{97.0\%} & \uit{63.4\%} & \uit{83.3\%}          \\
		& \textbf{ContraNet}\tnote{$\dagger$} & 79.4\% & 95.7\% & 38.2\% & 78.2\%          \\
		& \textbf{ContraNet}\tnote{$\dagger$}~\textbf{+ATC} & \textbf{93.7\%} & \textbf{99.3\%} & \textbf{89.7\%} & \textbf{95.1\%}\\
		& Trapdoor & 0.2\%         & 49.5\%          & 0.4\%           & 37.2\%          \\
		& DLA'20      & 17.0\%          & 55.9\%          & 0\%             & 13.5\%          \\
		& SID'21      & 8.9\%         & 50.9\%          & 0\%             & 11.4\%          \\
  	\hlinew{1pt}
	\end{tabular}
	\begin{tablenotes}
		\footnotesize
		\item[*] The adaptive objective function contains all losses, $\mathcal{L}_{ContraNet_{1}}$.

		\item[$\dagger$] The adaptive objective function only contains $\mathcal{L}_{DMM}$ (the weakest component of ContraNet under adaptive attack).
		\item[$\ddagger$] Results for baselines are from \cite{bryniarski2021evading}. ATC we use here is Gowal2020Uncovering~\cite{gowal2020uncovering} from RobustBench~\cite{croce2020robustbench}
	\end{tablenotes}
	\end{threeparttable} 
	\end{table}

%% file: 9.limitation_and_futurework.tex
\section{Discussions}
\label{sec:limitation}
\noindent
\textbf{Computation and memory costs.} 
\contranet constructs several additional modules (e.g., the conditional generator and the deep metric model) aside from the target DNN model for AE detection, which inevitably incurs extra computation and memory costs. The relative cost of \contranet is mainly dependent on the size and computational requirement of the protected DNN model. Tab.~\ref{tab:inference_cost} presents the comparison between \contranet and several classifiers used for \cifar dataset.  As can be seen, in comparison to SOTA classifiers, especially those using adversarial training techniques~\cite{croce2020robustbench}, \contranet has relatively low storage, \textit{Params},  and compute costs, \textit{FLOPs}.
Considering the trend in deep learning is to use an ever-larger pre-trained model for performance enhancement and the severity of AEs for safety-critical applications, we believe the extra computation and memory costs of \contranet are acceptable. 

\begin{figure}[t]
  \centering
  \includegraphics[width=0.85\linewidth]{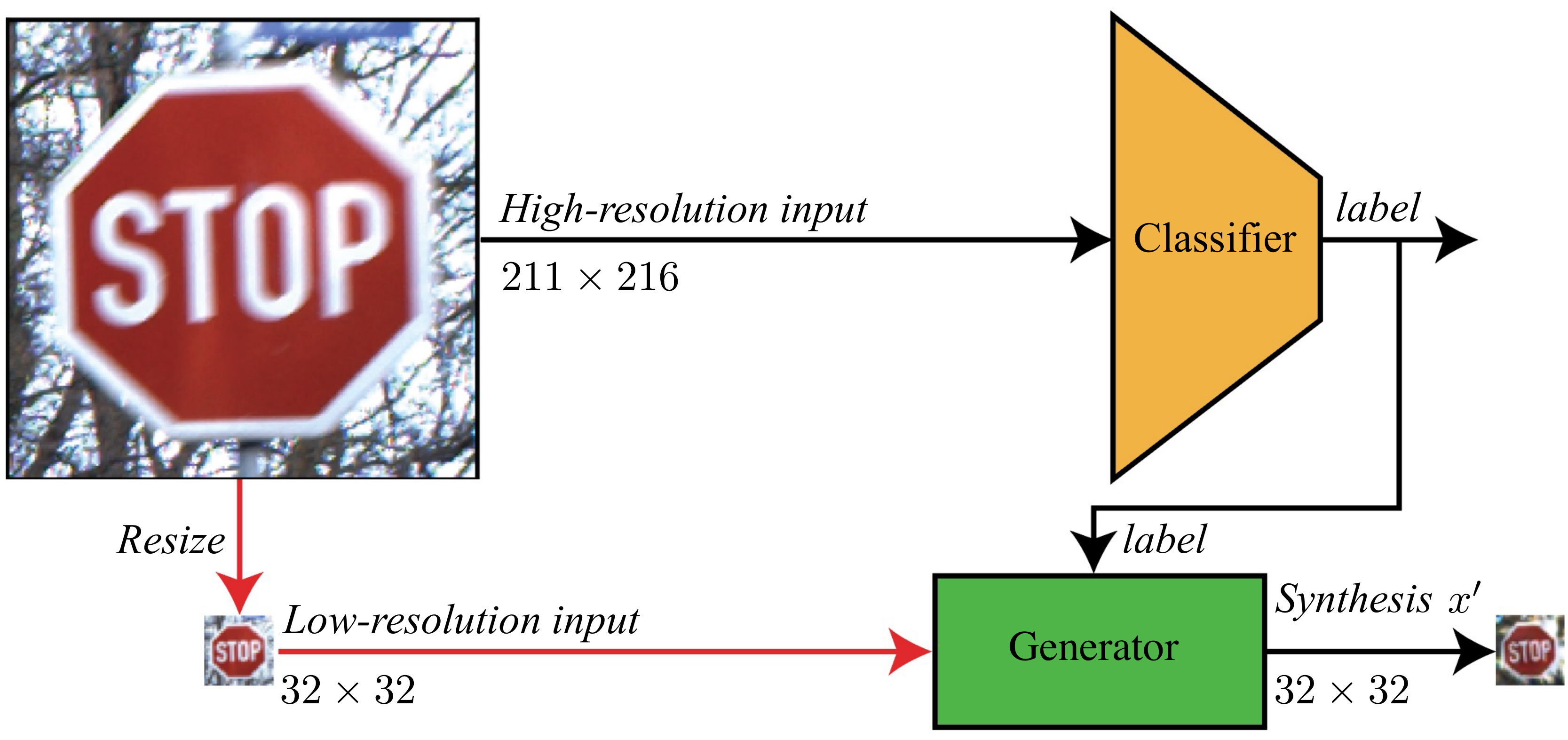}
  \caption{ Workflow of \contranet protecting the high-resolution classifier.    
  ContraNet can be performed on compressed inputs for a lower computational cost yet effective detection performance.}
  \label{fig:high-resolution}
  \vspace{-10pt}
\end{figure}

\begin{table}[t]\centering
  \caption{Computation and Storage Cost of ContraNet}
  \label{tab:inference_cost}
\begin{threeparttable}
  \setlength{\tabcolsep}{6mm}{
  \begin{tabular}{ccl}
    \hlinew{1pt}
     Modules & Params(M) & FLOPs(G) \\
    \hline
    DenseNet169~\cite{huang2017densely} & 12.49 & 0.27 \\
    Resnet152~\cite{huang2017densely} & 60.19 & 11.58 \\
    WideResNet-70-16\tnote{*}~\cite{zagoruyko2016wide} & 266.80 & 38.78 \\
    WideResNet-28-10\tnote{$\dagger$}~\cite{zagoruyko2016wide} & 36.48  & 5.25 \\
    \hline
    Encoder & 0.99 & 0.02 \\
    Generator & 9.42 & 3.83 \\
    Deep Metric Model & 5.82 & 0.18 \\
    Discriminator & 2.41 & 0.62 \\
    \hline
    Total ContraNet & 10.35 & 4.65 \\ 
    \hlinew{1pt}
  \end{tabular}}
  \begin{tablenotes}
    \footnotesize
    \item[*] used by Gowal2020Uncovering in Sec.~\ref{sec:ContraNet against AA attacks} and \ref{sec:Adaptive} and Rebuffi2021Fixing\_70 in Sec.~\ref{sec:ContraNet against AA attacks}, from \cite{croce2020robustbench}.
    \item[$\dagger$] used by Rebuffi2021Fixing\_28 and Sehwag2020Hydra in Sec.~\ref{sec:ContraNet against AA attacks}, from \cite{croce2020robustbench}.
  \end{tablenotes}
\end{threeparttable}
\vspace{-10pt}
\end{table}

\noindent
\textbf{Dealing with high-resolution images.}
In our earlier experiments, most of the images in our datasets are  low-resolution. 
To evaluate \contranet's scalability on high-resolution (HR) images, we construct a new dataset \imagenetten, a subset of the \imagenet~\cite{deng2009imagenet}.
We combine the classes with similar semantic information in \imagenet to be a general class, ending up ten different classes with 5000 images for each class, the same number as \cifar. 
Containing more data in each class can mitigate the common overfitting issue~\cite{zhao2020differentiable}.
We use $256 \times 256$ as the input size.
For each class, we randomly pick 4800 images for training, 200 images for testing.
As for the classifier, we choose ResNet50, which achieves 99\% accuracy on the test set. 
\contranet trained on \imagenetten achieves $Acc_{dec}=95\%$ on clean samples (FPR$=5\%$), which is acceptable compared to the classifier's original accuracy.
We summarize \contranet's performance against four white-box attacks in Tab.~\ref{tab:high-resolution10}. 

Comparing to the performance on low-resolution images (refer to Tab.~\ref{tab:robustacc}-~\ref{tab:TPR}), \contranet achieves better results on HR images, thanks to the high quality of images and correspondingly scaled-up model capability.
We visualize HR images with their synthesis in Appendix~\ref{appendix:more_synthesis}.    

\begin{table}[t]
  \centering

  \caption{The performance of \contranet on \imagenetten.}
  \label{tab:high-resolution10}
  \setlength{\tabcolsep}{3mm}{
  \begin{tabular}{lcccc}
  \hlinew{1pt}
  Attack Method         & \bm{$Acc_{rob}$}              & \textbf{TPR@FPR5\%}                   \\ \hline
  $PGD_{{\ell}_{\infty}}$           & 93.85\%          & 94.19\%           \\
  $BIM_{{\ell}_{\infty}}$           & 92.15\%          & 92.63\%                 \\
  $EDA_{{\ell}_{1}}$                & 95.8\%           & 96.16\%                \\
  $C\&W_{{\ell}_{\infty}}$          & 95.8\%           & 96.11\%                \\ 
  \bottomrule
  \end{tabular}}
\vspace{-8pt}
\end{table}

\begin{table}[t]
  \centering

  \caption{\contranet protects classifier with high-resolution inputs.}
  \label{tab:high-resolution}
  \setlength{\tabcolsep}{2mm}{
  \begin{tabular}{lcccc}
  \hlinew{1pt}
  Dataset       & \multicolumn{2}{c}{\imagenet-\cinic} & \multicolumn{2}{c}{\gtsrb-\btsc} \\ \hline
  Attack Method & \bm{$Acc_{rob}$}              & \textbf{TPR@FPR5\%}             & \bm{$Acc_{rob}$}          & \textbf{TPR@FPR5\%}           \\ \hline
  $PGD_{{\ell}_{\infty}}$           & 86.30\%          & 87.91\%         & 92.86\%        & 95.81\%       \\
  $BIM_{{\ell}_{\infty}}$           & 85.40\%          & 86.85\%         & 93.28\%        & 95.51\%       \\
  $EDA_{{\ell}_{1}}$                & 85.51\%          & 87.06\%         & 92.86\%        & 96.57\%       \\
  $C\&W_{{\ell}_{\infty}}$          & 86.20\%          & 87.38\%         & 93.28\%        & 97.06\%       \\
  \bottomrule
  \end{tabular}}
\vspace{-8pt}
\end{table}

At the same time, the computational overhead might be a concern when applying \contranet to HR images directly.
Note that \contranet detects AEs based on semantic contradiction, while the semantic information is largely preserved after compression.
Therefore, a possible solution to reduce the overhead is to resize the input images to compressed ones before feeding them to \contranet.
Whereas the classifier can remain unchanged.
Fig.~\ref{fig:high-resolution} illustrates this workflow.

We verify the feasibility of this solution by transferring \contranet trained on \cifar or \gtsrb to the new datasets (\imagenet-\cinic, \gtsrb-\btsc) with high-resolution images and test their performance.
For \imagenet-\cinic, we construct the dataset following the selection procedures of \cinic dataset~\cite{darlow2018cinic10}~\footnote{\cinic~\cite{darlow2018cinic10} provides a class mapping from \cifar to \imagenet. We follow the mapping and fetch HR images from \imagenet.}.
\imagenet-\cinic shares similar classes as \cifar while contains images of higher resolutions.
Then, we train a ResNet152 classifier on \imagenet-\cinic with $224\times 224$ as the input size.
After that, we apply \contranet trained on $32 \times 32$ to protect the classifier.
The workflow for detection is shown in Fig.~\ref{fig:high-resolution}.
We directly employ the \E and \G of \contranet trained on \cifar ($32 \times 32$) and fine-tune the deep metric model on \imagenet-\cinic for 15 epochs.
For \gtsrb-\btsc, since \gtsrb dataset already contains partial HR images, we train a classifier with input size of $224 \times 224$ directly.
We only boost the \emph{test} dataset with extra HR traffic sign images from \textit{Belgium Traffic Sign}~\cite{6707049}.
As can be seen in Tab.~\ref{tab:high-resolution}, both the $Acc_{rob}$ and TPR@FPR5\% are comparable to the results on \cifar or \gtsrb dataset (refer to Tab.~\ref{tab:robustacc}-\ref{tab:TPR}).
Such generalization capability proves it possible for \contranet to detect AEs on resized images. 

To better understand the advantage of image compression, we compare the overhead of \contranet on different input sizes.
The computational cost (FLOPs) and storage consumption (Params) for \contranet with $32 \times 32$ inputs are only 4.25\% and 10.16\% to those on $256 \times 256$~\footnote{cGAN typically works with an integral power of 2 as the input size.} inputs. 
Further, if compared with the ResNet152 on $224 \times 224$ inputs, the numbers will change to 31.99\% and 40.63\%, respectively.
Therefore, image compression effectively reduces the overhead of \contranet without lowering AE detection capabilities.

\noindent\textbf{Data pre-processing attack.}
There is a kind of adversarial attack targeting at the \textit{data pre-processing} stage.
For example, the \textit{Image-Scaling} attacks can obtain an output image that resembles the target image after downscaling the input image~\cite{QuiKleArp20, xiao2019seeing}.
The image-scaling attacks are operated on the image resize step, which may alter the semantic information of the image fed to the DNN model.
Such impact on DNN system is preserved in all subsequent steps.

\contranet cannot detect such kinds of AEs because there is no semantic contradiction between the input of DNN and the discriminative features from the classifier.
However, the defender can easily fix the vulnerability caused by an image-scaling attack due to the explicitly working mechanism of the image-scaling function~\cite{xiao2019seeing}.
Such a defensive mechanism can work seamlessly with \contranet without affecting each other.

\noindent
\textbf{Limitations.} 
One of the limitations of ContraNet is that it has difficulty in differentiating classes that are inherently similar in semantics. For example, there are more than one hundred types of dogs in ImageNet dataset~\cite{deng2009imagenet}, and the synthetic images for some of them could be pretty similar. 

However, this is usually not a security concern. On the one hand, from the adversary's perspective, an AE attack targeted on a class similar to its original is usually of minor severity, e.g., from one type of dog to another type. On the other hand, if a particular class needs to be secured, its semantic meaning should be identifiable from others\footnote{Otherwise, it may be misclassified even without an attack, causing security concerns.}, e.g., the ``Stop'' sign. 

%% file: 8.conclusion_and_futureworks.tex
\section{Conclusion and Future Work}
\label{sec:conclusion}
We proposed a novel AE detection framework, ContraNet, which focuses on identifying the intrinsic contradiction between AE's semantic meaning (in human eyes) and its DNN-extracted discriminative features (in DNN's eyes). We empirically evidence the effectiveness of ContraNet via adequate white-box and adaptive attacks. Experimental results show that ContraNet outperforms state-of-the-art AE detectors by a large margin. We have also shown that ContraNet can be combined with adversarial training techniques to enhance DNN model robustness further.

While most AE attacks and defenses focus on image classifiers and ContraNet is also applied in this context, adversarial examples for other DNN models and tasks have proliferated recently. We believe the basic concept of ContraNet still holds, and we plan to extend it to detect these new types of AEs.

%% file: appendix.tex
\appendix

\subsection{White-box Attack Configurations}
\label{sec::attack_Configuration}
The hyperparameters we used for the four typical untargeted attacks are as follow.
For PGD and BIM, we set the iteration step to 50.
We set the adversarial perturbation for PGD and BIM to 8/255 for ${\ell}_{\infty}$ norm, and 1 for ${\ell}_{2}$ norm.
For C\&W and EAD, we set the binary search steps to 9, the step size to $0.01$, optimization iteration to 1000, and the attack confidence to 0.
For C\&W, we employ {\it Adam}\cite{Bengio2014Adam} as the optimizer.

\subsection{ROC Curves with AUC under ${\ell}_2$-norm} 
\label{appendix:roc}
\begin{figure}[h]
    \centering
    \includegraphics[width=0.8\linewidth]{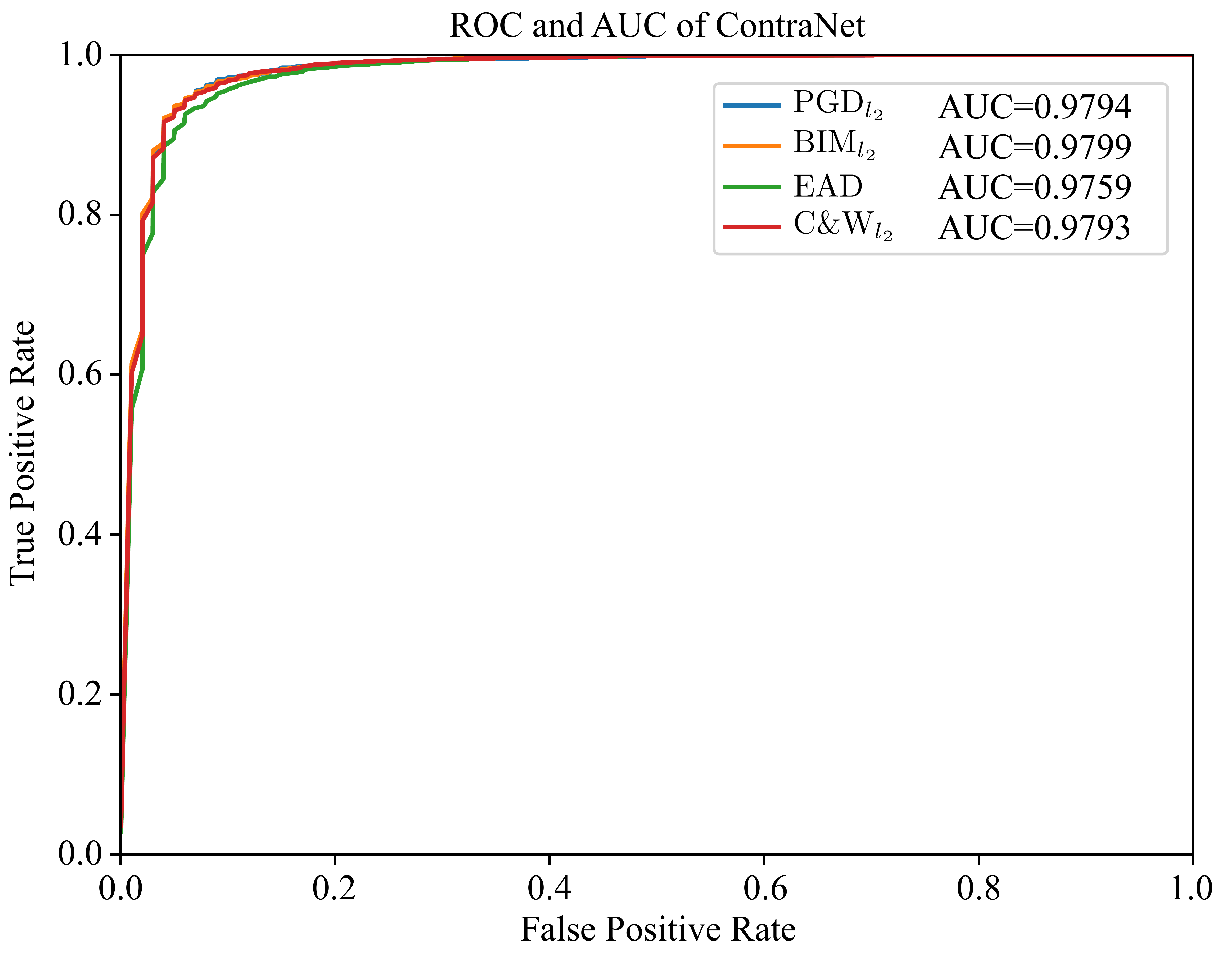}
    \caption{ROCs are generated under four attacks PGD$_{{\ell}_{2}}$, BIM$_{{\ell}_{2}}$, EAD, C\&W$_{{\ell}_{2}}$, along with corresponding AUC (higher is better) demonstrated in legend block.}
    \label{fig:rocl2}
    
\end{figure}

We generate the ROC curves together with AUC on \cifar under various attacks under ${\ell}_2$-norm, as shown in Fiugre~\ref{fig:rocl2}.  We can observe that the TPR can rapidly increase when the FPR is increased by a modest amount.
This demonstrates that ContraNet is capable of detecting AE with high accuracy while having a negligible influence on normal samples.
Additionally, AUC  
ContraNet's AUCs, which are used to indicate the overall performance for various attacks are all close to $1$, indicating that ContraNet performs admirably at AE detection.

\subsection{$Acc_{rob}$ under ${\ell}_2$-norm Attacks}
\label{appendix:l2_robust_acc}
Table~\ref{tab:robustaccL2} shows the $Acc_{rob}$ for ContraNet and every baseline under attacks with ${\ell}_{2}=1$ (for EAD, it is the EAD distance).  As can be observed, ContraNet's $Acc_{rob}$ is more stable than baselines and can maintain a high accuracy value across a variety of attack types and datasets. ContraNet's detection mechanism accounts for the above outperformed attack-agnostic feature.
ContraNet makes use of semantic contradiction, which requires no knowledge of the secured model and makes no assumptions about the attack methods.
As a result, ContraNet achieves a higher degree of generality when confronted with a variety of attacks.
\useunder{\uline}{\ul}{}
\begin{table*}[htbp]
	\caption{$Acc_{rob}$ under White-box Attack}
	\label{tab:robustaccL2}
	\centering
	\begin{threeparttable}
	\setlength{\tabcolsep}{5mm}{
	\begin{tabular}{@{}llcccccc@{}}
	\toprule
	\multicolumn{2}{c}{Dataset}   & \cifar         & \multicolumn{1}{c}{\gtsrb} & \multicolumn{1}{c}{\mnist} & \cifar  & \multicolumn{1}{c}{\gtsrb} & \multicolumn{1}{c}{\mnist}\\ \midrule
	Attack Method & Defense            & \multicolumn{3}{c}{\bm{$Acc_{rob}$}\textbf{@FPR5\%}$\uparrow$  }   & \multicolumn{3}{c}{\bm{$Acc_{rob}$}\textbf{@FPR3\%}$\uparrow$  } \\ \midrule
	\multirow{5}{*}{$PGD_{{\ell}_{2}}$}  
	              &\textbf{ContraNet}  & \textbf{90.23\%} & \uit{94.63\%}    & \textbf{99.96\%} & \textbf{79.58\%}       & {\ul \textit{89.52\%}}    & \textbf{99.92\%} \\
				  & Trapdoor           & 64.47\%          & \textbf{95.07\%} & \uit{99.79\%}   & 61.43\%                & \textbf{94.41\%}          & {\ul \textit{99.79\%}}       \\
				  & RR                 & 39.76\%          & 34.42\%          & 96.22\%       & 37.66\%                & 33.59\%                   & 94.24\%         \\
				  & MagNet             & \uit{73.18\%}    & 39.77\%          & 97.23\%       & {\ul \textit{71.69\%}} & 36.65\%                   & 96.77\%    \\
				  & FS                 & 64.61\%          & 86.56\%          & 94.78\%        & 58.40\%                & 80.07\%                   & 94.50\%         \\ \hline
	\multirow{5}{*}{$BIM_{{\ell}_{2}}$} 
				  & \textbf{ContraNet} & \textbf{90.47\%} & \uit{94.72\%}    & \textbf{99.96\%}   & \textbf{79.79\%}       & {\ul \textit{90.18\%}}    & \textbf{99.88\%} \\
				  & Trapdoor           & \uit{73.56\%}    & \textbf{96.71\%} & \uit{99.55\%}      & {\ul \textit{71.55\%}} & \textbf{96.05\%}          & {\ul \textit{99.51\%}}     \\
				  & RR                 & 32.81\%          & 24.79\%          & 95.07\%       & 30.96\%                & 24.30\%                   & 92.80\%    \\
				  & MagNet             & 59.19\%          & 39.56\%          & 96.40\%           & 57.74\%                & 35.82\%                   & 96.03\%    \\
				  & FS                 & 49.12\%          & 92.05\%          & 94.78\%         & 43.60\%                & 86.61\%                   & 94.33\%   \\ \hline
	\multirow{5}{*}{$EAD$} 
				  & \textbf{ContraNet} & \textbf{86.55\%} & \textbf{98.51\%} & \uit{96.15\%}          & \textbf{74.34\%}       & \textbf{96.32\%}          & {\ul \textit{95.16\%}}          \\
				  & Trapdoor           & 43.32\%          & 6.53\%           & 85.35\%      & 40.45\%                & 0.40\%                    & 82.63\%       \\
				  & RR                 & 11.39\%          & 13.06\%          & \textbf{99.87\%}  & 7.27\%                 & 6.49\%                    & \textbf{99.68\%}\\
				  & MagNet             & \uit{73.59\% }   & \uit{97.89\% }   & 86.59\%        & {\ul \textit{71.65\%}} & {\ul \textit{95.49\%}}    & 84.73\%     \\
				  & FS                 & 50.17\%          & 44.91\%          & 5.84\%          & {\ul \textit{71.65\%}} & 34.81\%                   & 3.56\% \\ \hline
	\multirow{5}{*}{$C\&W_{{\ell}_{2}}$}  
	              & \textbf{ContraNet} & \textbf{89.49\%} & \textbf{99.00\%} & \uit{99.42\%}      & {\ul \textit{78.10\%}} & \textbf{97.38\%}          & {\ul \textit{99.46\%}}     \\
				  & Trapdoor           & 30.38\%          & 3.00\%           & 37.41\%      & 28.52\%                & 0.10\%                    & 34.78\%     \\
				  & RR                 & 11.47\%          & 14.54\%          & \textbf{99.87\%} & 7.30\%                 & 8.65\%                    & \textbf{99.71\%}    \\
				  & MagNet             & \uit{85.43\%}    & \uit{98.09\%}    & 73.34\%         & \textbf{84.19\%}       & {\ul \textit{96.55\%}}    & 71.27\%   \\
				  & FS                 & 36.88\%          & 44.3\%           & 8.15\%         & 26.12\%                & 33.15\%                   & 5.17\%          \\ 
	 
				  \bottomrule
	\end{tabular}}
	\begin{tablenotes}
		\footnotesize
		\item[$\diamond $] The \textbf{bolded} values are the highest performance. The \uit{underlined italicized} values are the second highest performance.
	\end{tablenotes}
	\end{threeparttable}
	\end{table*}

\subsection{Detector's Metric under $l_2$-norm Attacks}
\label{appendix:l2_detector_acc}
Table~\ref{tab:TPRL2} shows the TPR@FPR 5\% for ContraNet and every baseline under attacks with ${\ell}_{2}=1$ (for EAD, it is the EAD distance).
As can be seen in Table~\ref{tab:TPRL2}, ContraNet outperforms, or at the very least performs on par with other detection-based defenses.
In particular, while dealing with the \cifar dataset, whose distribution is rather complex, ContraNet outperforms all other approaches by a significant margin in all attacks.
\begin{table*}[htbp]
	\caption{TPR@FPR 5\% or 3\% of White-box Attack}
	\label{tab:TPRL2}
	\centering
	\begin{threeparttable}
	\setlength{\tabcolsep}{5mm}{
	\begin{tabular}{llcccccc}
	\toprule
	\multicolumn{2}{c}{Dataset}   & \cifar         & \multicolumn{1}{c}{\gtsrb} & \multicolumn{1}{c}{\mnist} & \cifar         & \multicolumn{1}{c}{\gtsrb} & \multicolumn{1}{c}{\mnist}\\ \midrule
	Attack Method & Defense              & \multicolumn{3}{c}{\textbf{TPR@FPR5\%}}   & \multicolumn{3}{c}{\textbf{TPR@FPR3\%}}              \\ \midrule
	\multirow{5}{*}{$PGD_{{\ell}_{2}}$}  
					& \textbf{ContraNet} & \textbf{92.42\%} & \uit{94.62\%}    & \textbf{100\%}   & \textbf{81.96\%}       & {\ul \textit{89.83\%}}    & \textbf{100\%}\\
					& Trapdoor           & 68.66\%          & \textbf{95.08\%} & \uit{99.83\%}     & 65.96\%                & \textbf{94.50\%}          & {\ul \textit{99.83\%}}     \\
					& RR                 & 10.12\%          & 2.07\%           & 81.19\%          & 6.30\%                 & 0.86\%                    & 68.34\%  \\
					& MagNet             & \uit{76.15\%}    & 36.48\%          & 61.17\%          & {\ul \textit{74.82\%}} & 33.18\%                   & 53.40\% \\
					& FS                 & 66.98\%          & 85.58\%          & 9.71\%           & 60.74\%                & 78.66\%                   & 4.85\%  \\ \hline
	\multirow{5}{*}{$BIM_{{\ell}_{2}}$}  
					& \textbf{ContraNet} & \textbf{92.76\%} & 94.92\%          & \textbf{100\%}   & \textbf{82.13\%}       & {\ul \textit{90.41\%}}    & \textbf{100\%}\\
					& Trapdoor           & \uit{76.79\%}    & \textbf{96.81\%} & \uit{99.54\%}     & {\ul \textit{74.89\%}} & \textbf{96.18\%}          & {\ul \textit{99.50\%}}     \\
					& RR                 & 7.9\%            & 1.54\%           & 77.45\%          & 5.00\%                 & 0.98\%                    & 64.19\% \\
					& MagNet             & 61.67\%          & 37.67\%          & 50.91\%          & 60.47\%                & 33.78\%                   & 45.45\% \\
					& FS                 & 51.26\%          & 91.96\%          & 9.09\%            & 45.64\%                & 85.94\%                   & 6.36\%    \\ \hline
	\multirow{5}{*}{$EAD$}  
					& \textbf{ContraNet} & \textbf{89.12\%} & \textbf{98.68\%} & \uit{96.26\%}      & \textbf{77.22\%}       & \textbf{96.79\%}          & {\ul \textit{95.25\%}}                \\
					& Trapdoor           & 48.85\%          & 6.69\%           & 86.03\%          & 31.69\%                & 0.39\%                    & 83.34\%   \\
					& RR                 & 10.07\%          & 10.9\%           & \textbf{99.95\%}  & 6.53\%                 & 4.82\%                    & \textbf{99.97\%}\\
					& MagNet             & \uit{76.65\%}    & \uit{98.35\%}    & 87.65\%           & {\ul \textit{74.98\%}} & {\ul \textit{96.58\%}}    & 85.76\% \\
					& FS                 & 51.98\%          & 45.38\%          & 5.88\%          & {\ul \textit{74.98\%}} & 55.34\%                   & 3.61\%    \\ \hline
	\multirow{5}{*}{$C\&W_{{\ell}_{2}}$}   
					& \textbf{ContraNet} & \textbf{92.25\%} & \textbf{99.11\%}  & \uit{99.7\% }     & {\ul \textit{81.23\%}} & \textbf{97.84\%}          & {\ul \textit{99.62\%}}              \\
					& Trapdoor           & 33.67\%          & 0.28\%            & 31.02\%           & 45.67\%                & 0.07\%                    & 28.15\%\\
					& RR                 & 10.17\%          & 12.54\%           & \textbf{100\%}    & 6.45\%                 & 7.19\%                    & \textbf{100\%}   \\
					& MagNet             & \uit{89.25\%}    & \uit{98.56\%}     & 73.75\%          & \textbf{88.33\%}       & {\ul \textit{97.67\%}}    & 71.61\%\\
					& FS                 & 57.84 \%         & 44.76\%           & 6.63\%          & 27.05\%                & 33.43\%                   & 3.59\%  \\ 
								
		\bottomrule
	\end{tabular}}
	\begin{tablenotes}
		\footnotesize
		\item[$\diamond $] The \textbf{bolded} values are the highest performance. The \uit{underlined italicized} values are the second highest performance.
	\end{tablenotes}
	\end{threeparttable}
\end{table*}

\subsection{Model architectures}
\begin{figure}[htbp]
    \centering
    \includegraphics[width=\linewidth]{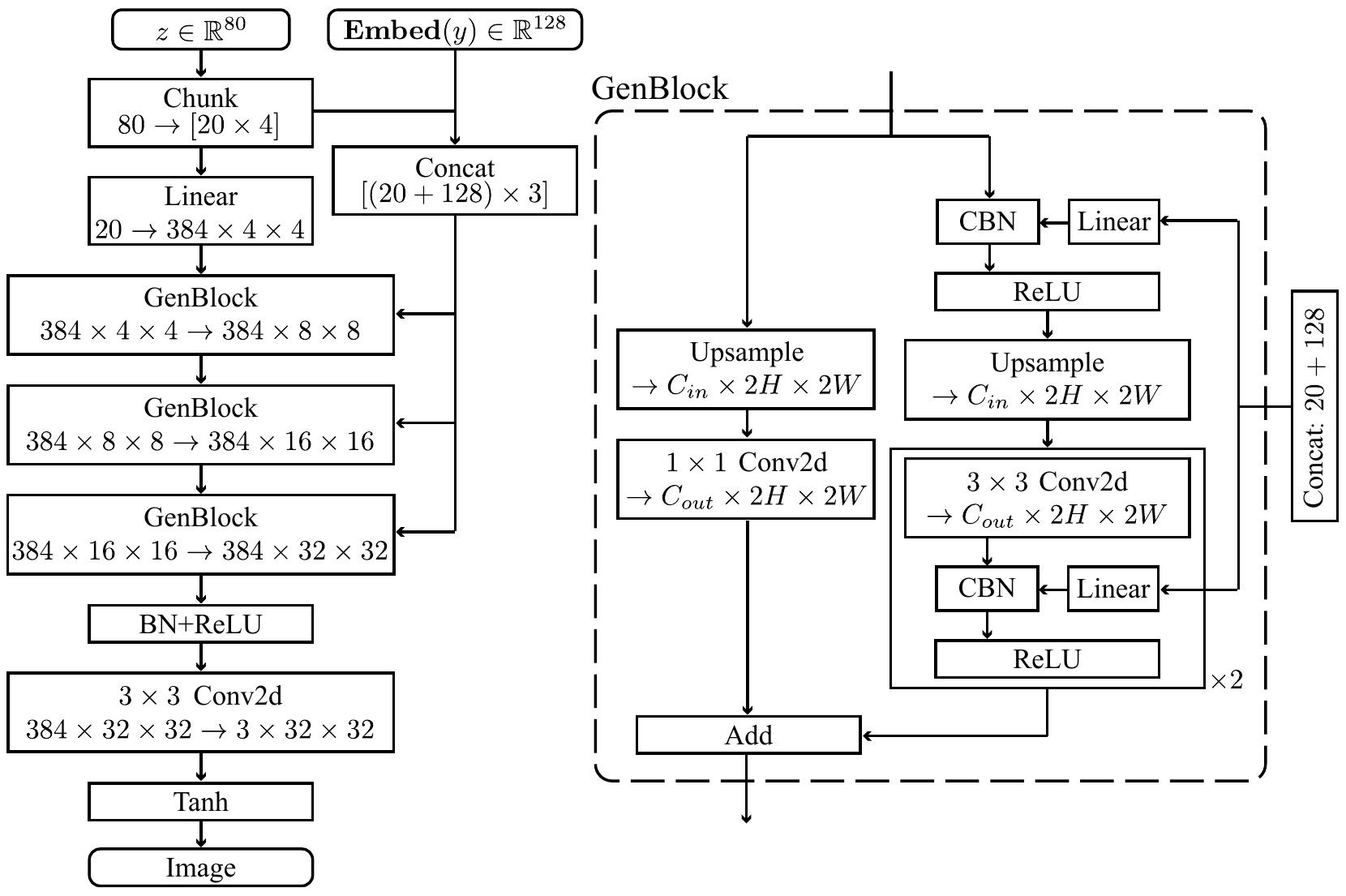}
    \caption{Generator's Architecture of \contranet.}
    \label{fig:generator_architecture}
    
\end{figure}

\begin{figure}[htbp]
    \centering
    \includegraphics[width=\linewidth]{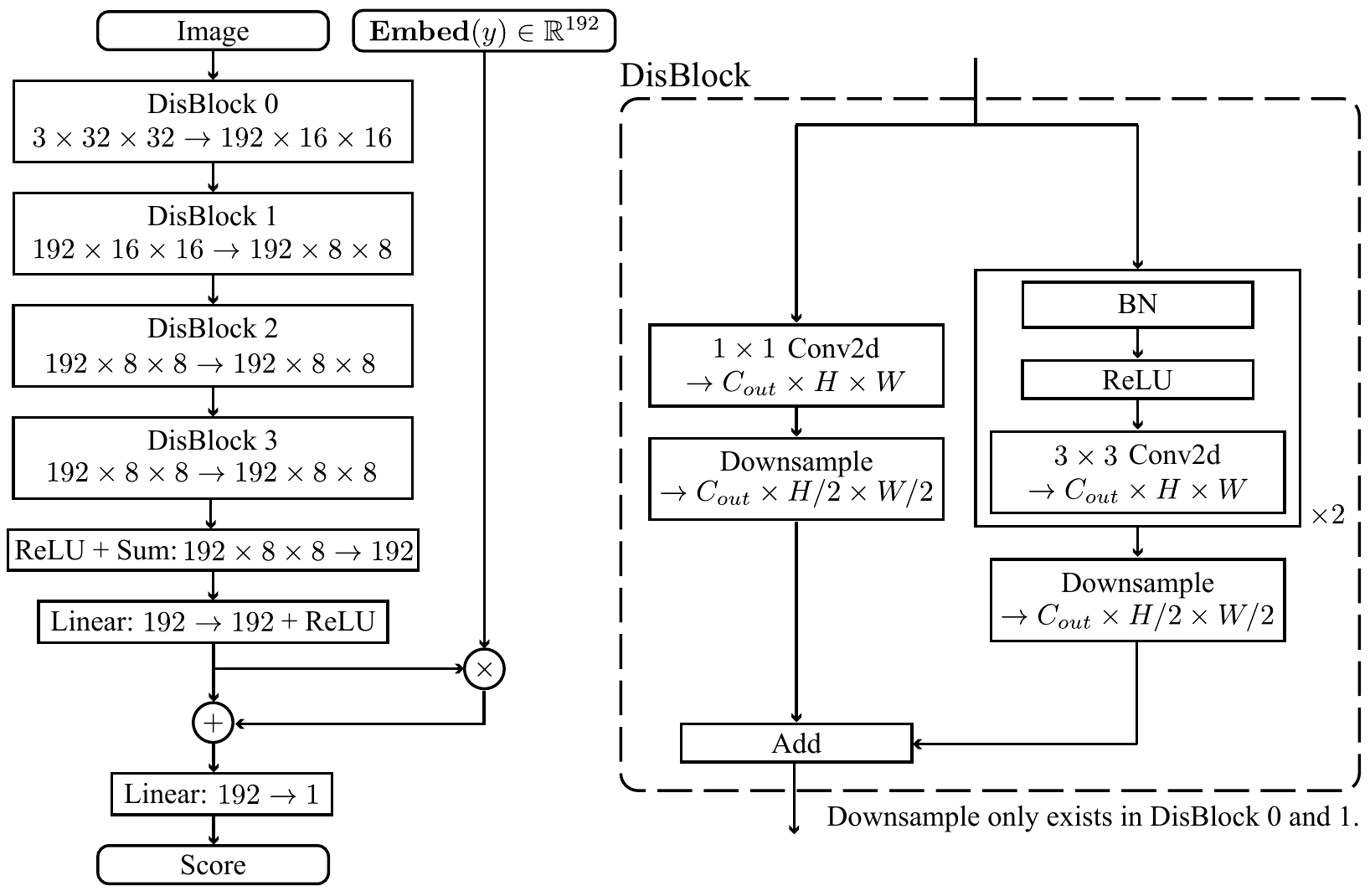}
    \caption{Discriminator's Architecture of \contranet.}
    \label{fig:discriminator_architecture}
    
\end{figure}

In this section, we give a detailed description of the model architecture we used when implementing \contranet.
The architecture of the Generator is shown in Fig.~\ref{fig:generator_architecture}.
The architecture of the Discriminator is shown in Fig.~\ref{fig:discriminator_architecture}.
The architecture of Encoder is only convolutional neural networks.
We show the parameters for each layer of the Encoder and MLP in Tab.~\ref{tab:encoderarch} and Tab.~\ref{tab:MLParch}.

\label{appendix:architectures}

\begin{table}[ht]
	\caption{Parameters for Each Layer of the Encoder}
	\label{tab:encoderarch}
	\centering
	\begin{threeparttable}	
	\setlength{\tabcolsep}{3mm}{
			\begin{tabular}{@{}rcl@{}}
				\toprule
				Input shape     & \multicolumn{2}{c}{$32\times 32\times 3$}  \\ \midrule
				Operator  & parameters                       & Output shape           \\ \midrule
				Conv2d    & channel: $3\to 64$               & $16\times 16\times 64$ \\
				ReLU      & -                                & $16\times 16\times 64$ \\
				Conv2d    & channel: $64\to 128$             & $8\times 8\times 128$  \\
				BN + ReLU & -                                & $8\times 8\times 128$  \\
				Conv2d    & channel: $128\to 256$            & $4\times 4\times 256$  \\
				BN + ReLU & -                                & $4\times 4\times 256$  \\
				Conv2d    & channel: $256\to 80$, no padding & $1\times 1\times 80$   \\
				Linear    & shape: $80\to80$                 & $80$                   \\ \bottomrule
		\end{tabular}}
		\begin{tablenotes}
			\footnotesize
			\item[$\diamond $] Conv2d use kernel size $=4$, stride $=2$, padding $=1$ as default.
		\end{tablenotes}
	\end{threeparttable}
\end{table}

\begin{table}[ht]
	\caption{Parameters for Each Layer of the MLP}
	\label{tab:MLParch}
	\centering
	\setlength{\tabcolsep}{3mm}{
	\begin{tabular}{@{}rc@{}}
		\toprule
		Input          & $\mathbf{\{en, ep\}} \in {\mathbb{R}}^{2560} $ \\ \midrule
		Operator       & Parameters                                     \\ \midrule
		Linear         & shape: $2560\to 512$                           \\
		Dropout + ReLU & drop rate: 0.1                                 \\
		Linear         & shape: $512\to 64$                             \\
		Dropout+ReLU   & drop rate: 0.1                                 \\
		Linear         & shape: $64\to 2$                               \\ \bottomrule
	\end{tabular}}
\end{table}

\subsection{More Synthetic Images}
\label{appendix:more_synthesis}
\begin{figure*}[t]
    \centering
    \includegraphics[width=\linewidth]{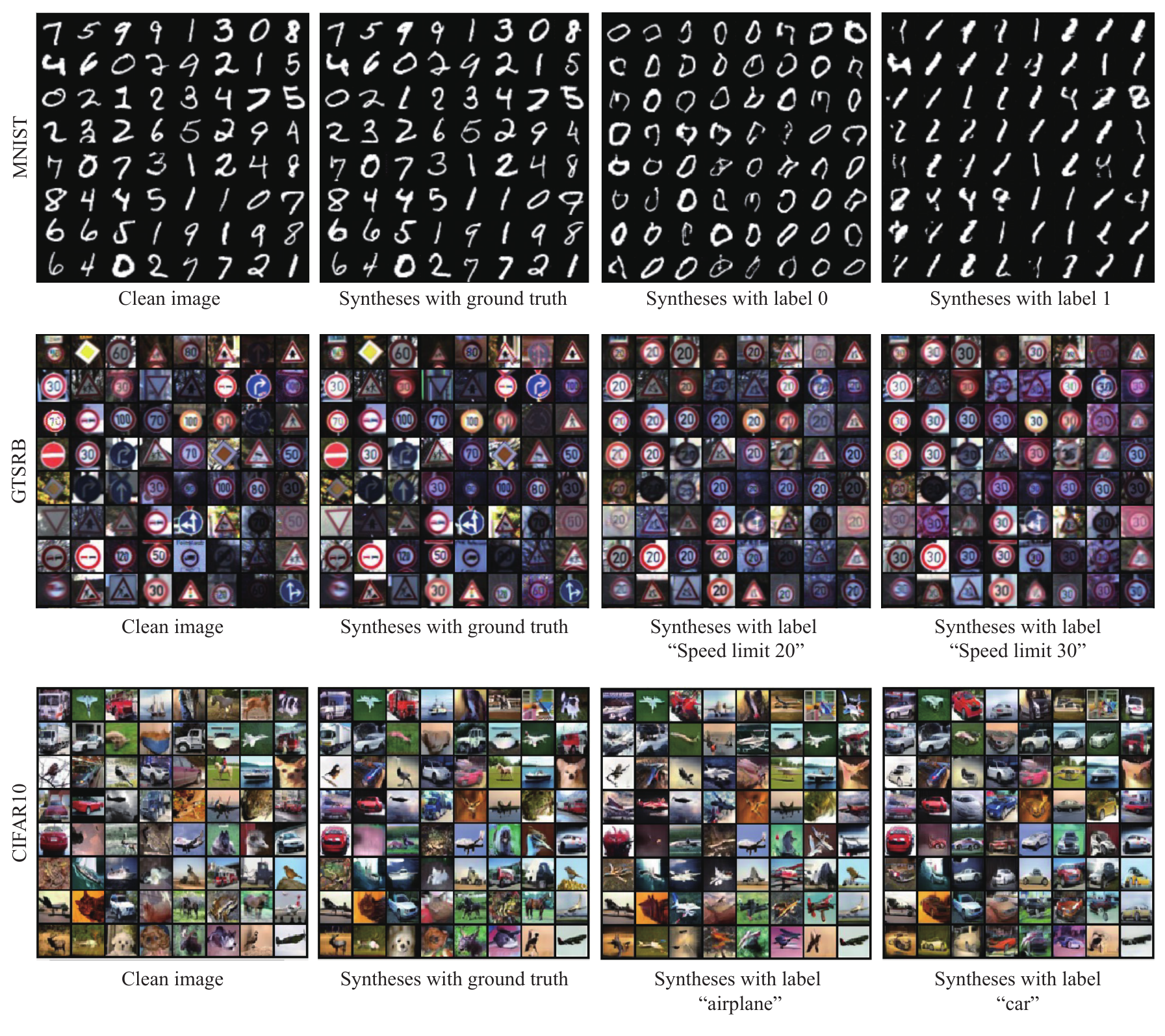}
    \caption{Synthetic images on \mnist, \gtsrb and \cifar. For each task we randomly sample 64 clean images from the test set, and generate the synthetic images under their groundtruth labels, the first and the second label in order. As can be observed, the semantic information contained in the synthetic images is significantly dependent on the given label.}
    \label{fig:more_fig}
    
\end{figure*}

\begin{figure*}[t]
    \centering
    \includegraphics[width=\linewidth]{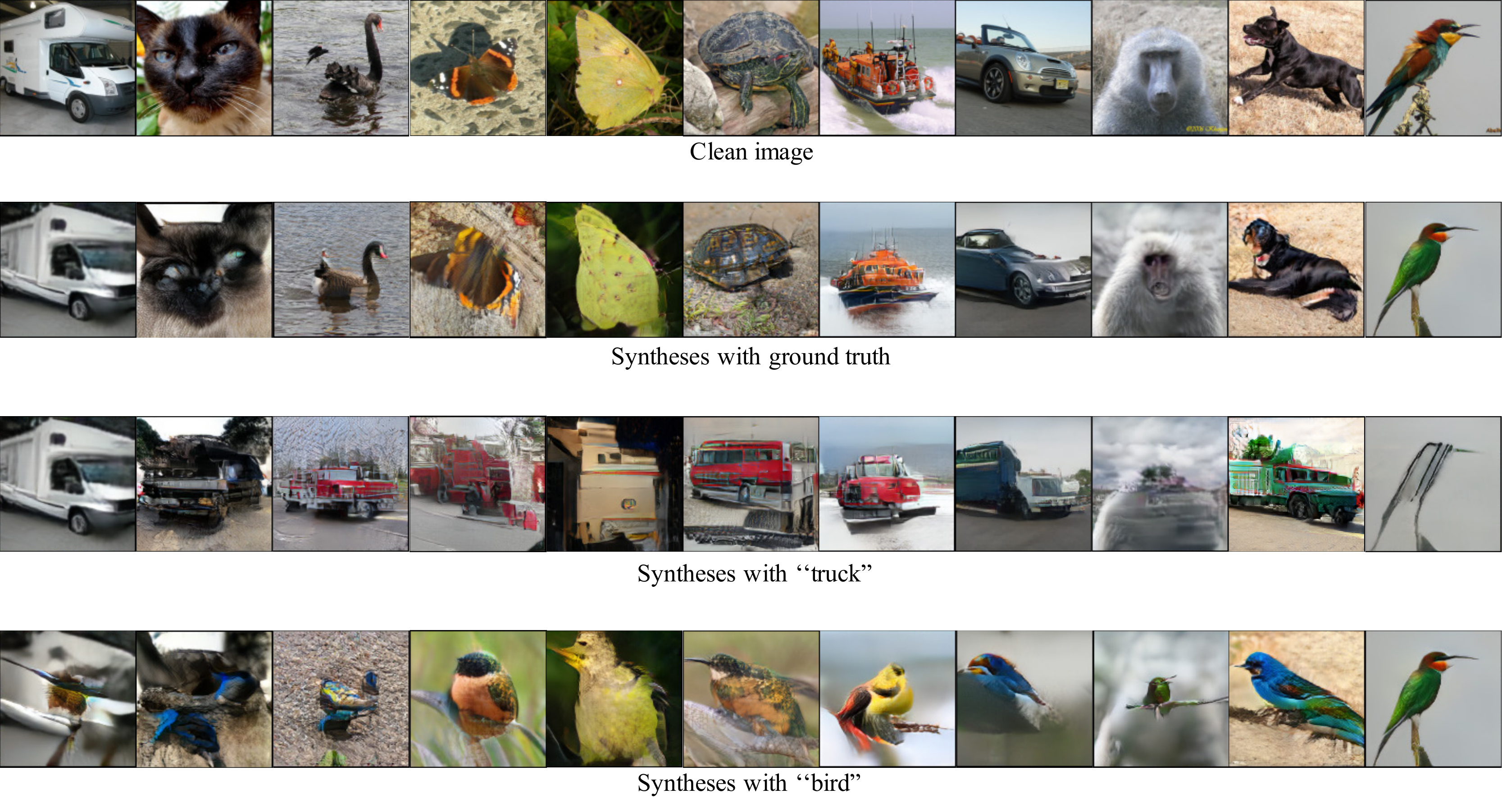}
    \caption{High-resolution synthetic images on \imagenetten. As can be observed, the semantic information contained in the synthetic images is significantly dependent on the given label.}
    \label{fig:imagenetten_fig}
    
\end{figure*}

We demonstrate empirically that the semantic information included in the synthetic images generated by ContraNet is strongly dependent on the associated labels.
Fig.~\ref{fig:more_fig} shows more synthetic images on \mnist, \gtsrb and \cifar.
Fig.~\ref{fig:imagenetten_fig} shows more synthetic images on \imagenetten.

%% file: main.bbl
\begin{thebibliography}{10}
\providecommand{\url}[1]{#1}
\csname url@samestyle\endcsname
\providecommand{\newblock}{\relax}
\providecommand{\bibinfo}[2]{#2}
\providecommand{\BIBentrySTDinterwordspacing}{\spaceskip=0pt\relax}
\providecommand{\BIBentryALTinterwordstretchfactor}{4}
\providecommand{\BIBentryALTinterwordspacing}{\spaceskip=\fontdimen2\font plus
\BIBentryALTinterwordstretchfactor\fontdimen3\font minus
  \fontdimen4\font\relax}
\providecommand{\BIBforeignlanguage}[2]{{%
\expandafter\ifx\csname l@#1\endcsname\relax
\typeout{** WARNING: IEEEtranS.bst: No hyphenation pattern has been}%
\typeout{** loaded for the language `#1'. Using the pattern for}%
\typeout{** the default language instead.}%
\else
\language=\csname l@#1\endcsname
\fi
#2}}
\providecommand{\BIBdecl}{\relax}
\BIBdecl

\bibitem{athalye2018obfuscated}
A.~Athalye, N.~Carlini, and D.~Wagner, ``Obfuscated gradients give a false
  sense of security: Circumventing defenses to adversarial examples,'' in
  \emph{International Conference on Machine Learning}.\hskip 1em plus 0.5em
  minus 0.4em\relax PMLR, 2018.

\bibitem{biggio2013evasion}
B.~Biggio, I.~Corona, D.~Maiorca, B.~Nelson, N.~{\v{S}}rndi{\'c}, P.~Laskov,
  G.~Giacinto, and F.~Roli, ``Evasion attacks against machine learning at test
  time,'' in \emph{Joint European conference on machine learning and knowledge
  discovery in databases}.\hskip 1em plus 0.5em minus 0.4em\relax Springer,
  2013, pp. 387--402.

\bibitem{board2005stochastic}
A.~Board, \emph{Stochastic modelling and applied probability}.\hskip 1em plus
  0.5em minus 0.4em\relax Springer, 2005.

\bibitem{breier2018practical}
J.~Breier, X.~Hou, D.~Jap, L.~Ma, S.~Bhasin, and Y.~Liu, ``Practical fault
  attack on deep neural networks,'' in \emph{ACM SIGSAC Conference on Computer
  and Communications Security}, 2018, pp. 2204--2206.

\bibitem{brock2018large}
A.~Brock, J.~Donahue, and K.~Simonyan, ``Large scale {GAN} training for high
  fidelity natural image synthesis,'' in \emph{International Conference on
  Learning Representations}, 2019.

\bibitem{bromley1993signature}
J.~Bromley, J.~W. Bentz, L.~Bottou, I.~Guyon, Y.~LeCun, C.~Moore,
  E.~S{\"a}ckinger, and R.~Shah, ``Signature verification using a “siamese”
  time delay neural network,'' \emph{International Journal of Pattern
  Recognition and Artificial Intelligence}, vol.~7, no.~04, pp. 669--688, 1993.

\bibitem{bryniarski2021evading}
O.~Bryniarski, N.~Hingun, P.~Pachuca, V.~Wang, and N.~Carlini, ``Evading
  adversarial example detection defenses with orthogonal projected gradient
  descent,'' \emph{arXiv preprint arXiv:2106.15023}, 2021.

\bibitem{buckman2018thermometer}
J.~Buckman, A.~Roy, C.~Raffel, and I.~Goodfellow, ``Thermometer encoding: One
  hot way to resist adversarial examples,'' in \emph{International Conference
  on Learning Representations}, 2018.

\bibitem{CarliniEvaluatingAdversarialRobustness2019}
N.~Carlini, A.~Athalye, N.~Papernot, W.~Brendel, J.~Rauber, D.~Tsipras, I.~J.
  Goodfellow, A.~Madry, and A.~Kurakin, ``On evaluating adversarial
  robustness,'' \emph{arXiv preprint arXiv:1902.06705}, 2019.

\bibitem{carlini2017adversarial}
N.~Carlini and D.~Wagner, ``Adversarial examples are not easily detected:
  Bypassing ten detection methods,'' in \emph{Proceedings of the 10th ACM
  Workshop on Artificial Intelligence and Security}, 2017, p. 3–14.

\bibitem{carlini2017magnet}
------, ``{MagnNet} and `efficient defenses against adversarial attacks' are
  not robust to adversarial examples,'' \emph{arXiv preprint arXiv:1711.08478},
  2017.

\bibitem{carlini2018towards}
------, ``Towards evaluating the robustness of neural networks,'' in \emph{2017
  IEEE Symposium on Security and Privacy}, 2017, pp. 39--57.

\bibitem{Carrara2018AdversarialED}
F.~Carrara, R.~Becarelli, R.~Caldelli, F.~Falchi, and G.~Amato, ``Adversarial
  examples detection in features distance spaces,'' in \emph{Proceedings of the
  European Conference on Computer Vision Workshops}, 2018.

\bibitem{chen2018ead}
P.-Y. Chen, Y.~Sharma, H.~Zhang, J.~Yi, and C.-J. Hsieh, ``{EAD}: Elastic-net
  attacks to deep neural networks via adversarial examples,'' in
  \emph{Thirty-second AAAI conference on artificial intelligence}, 2018.

\bibitem{Chen_2017}
P.-Y. Chen, H.~Zhang, Y.~Sharma, J.~Yi, and C.-J. Hsieh, ``Zoo: Zeroth order
  optimization based black-box attacks to deep neural networks without training
  substitute models,'' in \emph{Proceedings of the 10th ACM Workshop on
  Artificial Intelligence and Security}, 2017, p. 15–26.

\bibitem{chen2020adversarial}
T.~Chen, S.~Liu, S.~Chang, Y.~Cheng, L.~Amini, and Z.~Wang, ``Adversarial
  robustness: From self-supervised pre-training to fine-tuning,'' in
  \emph{Proceedings of the IEEE/CVF Conference on Computer Vision and Pattern
  Recognition}, 2020, pp. 699--708.

\bibitem{automobile}
M.~Cococcioni, F.~Rossi, E.~Ruffaldi, S.~Saponara, and B.~Dupont~de Dinechin,
  ``Novel arithmetics in deep neural networks signal processing for autonomous
  driving: Challenges and opportunities,'' \emph{IEEE Signal Processing
  Magazine}, pp. 97--110, 2021.

\bibitem{croce2020robustbench}
F.~Croce, M.~Andriushchenko, V.~Sehwag, E.~Debenedetti, N.~Flammarion,
  M.~Chiang, P.~Mittal, and M.~Hein, ``Robustbench: a standardized adversarial
  robustness benchmark,'' \emph{arXiv preprint arXiv:2010.09670}, 2020.

\bibitem{croce2020reliable}
F.~Croce and M.~Hein, ``Reliable evaluation of adversarial robustness with an
  ensemble of diverse parameter-free attacks,'' in \emph{International
  conference on machine learning}.\hskip 1em plus 0.5em minus 0.4em\relax PMLR,
  2020, pp. 2206--2216.

\bibitem{darlow2018cinic10}
L.~N. Darlow, E.~J. Crowley, A.~Antoniou, and A.~J. Storkey, ``Cinic-10 is not
  imagenet or cifar-10,'' 2018.

\bibitem{de2017modulating}
H.~de~Vries, F.~Strub, J.~Mary, H.~Larochelle, O.~Pietquin, and A.~Courville,
  ``Modulating early visual processing by language,'' in \emph{Conference on
  Neural Information Processing Systems}, 2017, pp. 1--14.

\bibitem{deng2009imagenet}
J.~Deng, W.~Dong, R.~Socher, L.-J. Li, K.~Li, and F.-F. Li, ``{ImageNet}: A
  large-scale hierarchical image database,'' in \emph{IEEE Conference on
  Computer Vision and Pattern Recognition}, 2009, pp. 248--255.

\bibitem{doersch2016tutorial}
C.~Doersch, ``Tutorial on variational autoencoders,'' \emph{arXiv preprint
  arXiv:1606.05908}, 2016.

\bibitem{dumoulin2016learned}
V.~Dumoulin, J.~Shlens, and M.~Kudlur, ``A learned representation for artistic
  style,'' in \emph{5th International Conference on Learning Representations},
  2017.

\bibitem{elfwing2018sigmoid}
S.~Elfwing, E.~Uchibe, and K.~Doya, ``Sigmoid-weighted linear units for neural
  network function approximation in reinforcement learning,'' \emph{Neural
  Networks}, vol. 107, pp. 3--11, 2018.

\bibitem{goodfellow2014explaining}
I.~J. Goodfellow, J.~Shlens, and C.~Szegedy, ``Explaining and harnessing
  adversarial examples,'' in \emph{3rd International Conference on Learning
  Representations}, 2015.

\bibitem{gowal2020uncovering}
S.~Gowal, C.~Qin, J.~Uesato, T.~Mann, and P.~Kohli, ``Uncovering the limits of
  adversarial training against norm-bounded adversarial examples,'' \emph{arXiv
  preprint arXiv:2010.03593}, 2020.

\bibitem{he2021feature}
C.~He, B.~B. Zhu, X.~Ma, H.~Jin, and S.~Hu, ``Feature-indistinguishable attack
  to circumvent trapdoor-enabled defense,'' in \emph{Proceedings of the 2021
  ACM SIGSAC Conference on Computer and Communications Security}, 2021, pp.
  3159--3176.

\bibitem{he2016deep}
K.~He, X.~Zhang, S.~Ren, and J.~Sun, ``Deep residual learning for image
  recognition,'' in \emph{Proceedings of the IEEE conference on computer vision
  and pattern recognition}, 2016, pp. 770--778.

\bibitem{HeAdversarialExampleDefenses2017}
W.~He, J.~Wei, X.~Chen, N.~Carlini, and D.~Song, ``Adversarial example
  defenses: Ensembles of weak defenses are not strong,'' in \emph{Proceedings
  of the 11th USENIX Conference on Offensive Technologies}, 2017.

\bibitem{he2020defending}
Z.~He, A.~S. Rakin, J.~Li, C.~Chakrabarti, and D.~Fan, ``Defending and
  harnessing the bit-flip based adversarial weight attack,'' in \emph{IEEE/CVF
  Conference on Computer Vision and Pattern Recognition}, 2020, pp.
  14\,095--14\,103.

\bibitem{hirano2021universal}
H.~Hirano, A.~Minagi, and K.~Takemoto, ``Universal adversarial attacks on deep
  neural networks for medical image classification,'' \emph{BMC medical
  imaging}, vol.~21, no.~1, pp. 1--13, 2021.

\bibitem{hoffer2018deep}
E.~Hoffer and N.~Ailon, ``Deep metric learning using triplet network,'' in
  \emph{International workshop on similarity-based pattern recognition}.\hskip
  1em plus 0.5em minus 0.4em\relax Springer, 2015, pp. 84--92.

\bibitem{huang2017densely}
G.~Huang, Z.~Liu, L.~Van Der~Maaten, and K.~Q. Weinberger, ``Densely connected
  convolutional networks,'' in \emph{Proceedings of the IEEE conference on
  computer vision and pattern recognition}, 2017, pp. 4700--4708.

\bibitem{kaissis2020secure}
G.~A. Kaissis, M.~R. Makowski, D.~R{\"u}ckert, and R.~F. Braren, ``Secure,
  privacy-preserving and federated machine learning in medical imaging,''
  \emph{Nature Machine Intelligence}, vol.~2, no.~6, pp. 305--311, 2020.

\bibitem{Bengio2014Adam}
D.~P. Kingma and J.~Ba, ``Adam: A method for stochastic optimization,'' in
  \emph{International Conference on Learning Representations}, 2015.

\bibitem{koch2015siamese}
G.~Koch, R.~Zemel, R.~Salakhutdinov \emph{et~al.}, ``Siamese neural networks
  for one-shot image recognition,'' in \emph{ICML deep learning workshop},
  vol.~2.\hskip 1em plus 0.5em minus 0.4em\relax Lille, 2015.

\bibitem{krizhevsky2009learning}
A.~Krizhevsky, ``Learning multiple layers of features from tiny images,''
  \emph{Master's thesis, University of Tront}, 2009.

\bibitem{krizhevsky2012imagenet}
A.~Krizhevsky, I.~Sutskever, and G.~E. Hinton, ``Imagenet classification with
  deep convolutional neural networks,'' \emph{Advances in neural information
  processing systems}, vol.~25, pp. 1097--1105, 2012.

\bibitem{kurakin2016bim}
A.~Kurakin, I.~J. Goodfellow, and S.~Bengio, ``Adversarial examples in the
  physical world,'' in \emph{International Conference on Learning
  Representations workshop}, 2017.

\bibitem{726791}
Y.~Lecun, L.~Bottou, Y.~Bengio, and P.~Haffner, ``Gradient-based learning
  applied to document recognition,'' \emph{Proceedings of the IEEE}, vol.~86,
  no.~11, pp. 2278--2324, 1998.

\bibitem{Lcuyer2019CertifiedRT}
M.~L{\'e}cuyer, V.~Atlidakis, R.~Geambasu, D.~Hsu, and S.~Jana, ``Certified
  robustness to adversarial examples with differential privacy,'' \emph{IEEE
  Symposium on Security and Privacy}, pp. 656--672, 2019.

\bibitem{li2019improving}
P.~Li, J.~Yi, B.~Zhou, and L.~Zhang, ``Improving the robustness of deep neural
  networks via adversarial training with triplet loss,'' in \emph{Proceedings
  of the Twenty-Eighth International Joint Conference on Artificial
  Intelligence, {IJCAI} 2019, Macao, China, August 10-16, 2019}, S.~Kraus,
  Ed.\hskip 1em plus 0.5em minus 0.4em\relax ijcai.org, 2019, pp. 2909--2915.

\bibitem{lin2020composite}
J.~Lin, L.~Xu, Y.~Liu, and X.~Zhang, ``Composite backdoor attack for deep
  neural network by mixing existing benign features,'' in \emph{Proceedings of
  the 2020 ACM SIGSAC Conference on Computer and Communications Security},
  2020, pp. 113--131.

\bibitem{ling2019deepsec}
X.~Ling, S.~Ji, J.~Zou, J.~Wang, C.~Wu, B.~Li, and T.~Wang, ``Deepsec: A
  uniform platform for security analysis of deep learning model,'' in
  \emph{2019 IEEE Symposium on Security and Privacy (SP)}, 2019, pp. 673--690.

\bibitem{liu2019detection}
J.~Liu, W.~Zhang, Y.~Zhang, D.~Hou, Y.~Liu, H.~Zha, and N.~Yu, ``Detection
  based defense against adversarial examples from the steganalysis point of
  view,'' in \emph{Proceedings of the IEEE/CVF Conference on Computer Vision
  and Pattern Recognition}, 2019, pp. 4825--4834.

\bibitem{8203770}
Y.~Liu, L.~Wei, B.~Luo, and Q.~Xu, ``Fault injection attack on deep neural
  network,'' in \emph{IEEE/ACM International Conference on Computer-Aided
  Design}, 2017, pp. 131--138.

\bibitem{robustml}
A.~Madry, A.~Athalye, D.~Tsipras, L.~Engstrom, D.~Wagner, N.~Carlini, P.~Liang,
  and Z.~Kolter, ``{RobustML},'' \url{https://www.robust-ml.org/}, accessed:
  2021-07-15.

\bibitem{madry-PGD}
A.~Madry, A.~Makelov, L.~Schmidt, D.~Tsipras, and A.~Vladu, ``Towards deep
  learning models resistant to adversarial attacks,'' in \emph{6th
  International Conference on Learning Representations}, 2018.

\bibitem{6707049}
M.~Mathias, R.~Timofte, R.~Benenson, and L.~Van~Gool, ``Traffic sign
  recognition — how far are we from the solution?'' in \emph{The
  International Joint Conference on Neural Networks (IJCNN)}, 2013.

\bibitem{MengMagNetTwoProngedDefense2017}
D.~Meng and H.~Chen, ``{{MagNet}}: {{A Two}}-{{Pronged Defense}} against
  {{Adversarial Examples}},'' in \emph{Proceedings of the 2017 {{ACM SIGSAC
  Conference}} on {{Computer}} and {{Communications Security}}}, 2017, pp.
  135--147.

\bibitem{MiyatoCGANsProjectionDiscriminator2018}
T.~Miyato and M.~Koyama, ``{{cGANs}} with {{Projection Discriminator}},'' in
  \emph{International {{Conference}} on {{Learning Representations}}}, 2018.

\bibitem{naren2021iomt}
N.~Naren, V.~Chamola, S.~Baitragunta, A.~Chintanpalli, P.~Mishra, S.~Yenuganti,
  and M.~Guizani, ``Iomt and dnn-enabled drone-assisted covid-19 screening and
  detection framework for rural areas,'' \emph{IEEE Internet of Things
  Magazine}, vol.~4, no.~2, pp. 4--9, 2021.

\bibitem{Pang2020ATO}
R.~Pang, H.~Shen, X.~Zhang, S.~Ji, Y.~Vorobeychik, X.~Luo, A.~X. Liu, and
  T.~Wang, ``A tale of evil twins: Adversarial inputs versus poisoned models,''
  \emph{ACM SIGSAC Conference on Computer and Communications Security}, 2020.

\bibitem{pmlr-v97-pang19a}
T.~Pang, K.~Xu, C.~Du, N.~Chen, and J.~Zhu, ``Improving adversarial robustness
  via promoting ensemble diversity,'' in \emph{Proceedings of the 36th
  International Conference on Machine Learning}, vol.~97.\hskip 1em plus 0.5em
  minus 0.4em\relax PMLR, 2019, pp. 4970--4979.

\bibitem{pang2021adversarial}
T.~Pang, H.~Zhang, D.~He, Y.~Dong, H.~Su, W.~Chen, J.~Zhu, and T.-Y. Liu,
  ``Adversarial training with rectified rejection,'' \emph{arXiv preprint
  arXiv:2105.14785}, 2021.

\bibitem{papernot2017practical}
N.~Papernot, P.~McDaniel, I.~Goodfellow, S.~Jha, Z.~B. Celik, and A.~Swami,
  ``Practical black-box attacks against machine learning,'' in
  \emph{Proceedings of the 2017 ACM on Asia conference on computer and
  communications security}, 2017, pp. 506--519.

\bibitem{papernot2016limitations}
N.~Papernot, P.~McDaniel, S.~Jha, M.~Fredrikson, Z.~B. Celik, and A.~Swami,
  ``The limitations of deep learning in adversarial settings,'' in \emph{2016
  IEEE European symposium on security and privacy (EuroS\&P)}, 2016, pp.
  372--387.

\bibitem{QuiKleArp20}
E.~Quiring, D.~Klein, D.~Arp, M.~Johns, and K.~Rieck, ``Adversarial
  preprocessing: Understanding and preventing image-scaling attacks in machine
  learning,'' in \emph{Proc. of USENIX Security Symposium}, 2020.

\bibitem{rebuffi2021fixing}
S.-A. Rebuffi, S.~Gowal, D.~A. Calian, F.~Stimberg, O.~Wiles, and T.~Mann,
  ``Fixing data augmentation to improve adversarial robustness,'' \emph{arXiv
  preprint arXiv:2103.01946}, 2021.

\bibitem{samangouei2018defense}
P.~Samangouei, M.~Kabkab, and R.~Chellappa, ``{Defense-GAN}: Protecting
  classifiers against adversarial attacks using generative models,'' in
  \emph{International Conference on Learning Representations}, 2018.

\bibitem{santurkar2019image}
S.~Santurkar, D.~Tsipras, B.~Tran, A.~Ilyas, L.~Engstrom, and A.~Madry, ``Image
  synthesis with a single (robust) classifier,'' in \emph{Proceedings of the
  33rd International Conference on Neural Information Processing Systems},
  2019, pp. 1262--1273.

\bibitem{sara2019image}
U.~Sara, M.~Akter, and M.~S. Uddin, ``Image quality assessment through fsim,
  ssim, mse and psnr—a comparative study,'' \emph{Journal of Computer and
  Communications}, vol.~7, no.~3, pp. 8--18, 2019.

\bibitem{schroff2015facenet}
F.~Schroff, D.~Kalenichenko, and J.~Philbin, ``Facenet: A unified embedding for
  face recognition and clustering,'' in \emph{Proceedings of the IEEE
  conference on computer vision and pattern recognition}, 2015, pp. 815--823.

\bibitem{Sehwag2020Hydra}
V.~Sehwag, S.~Wang, P.~Mittal, and S.~Jana, ``{HYDRA}: Pruning adversarially
  robust neural networks,'' in \emph{Conference on Neural Information
  Processing Systems}, 2020.

\bibitem{shafahi2019adversarial}
A.~Shafahi, M.~Najibi, A.~Ghiasi, Z.~Xu, J.~Dickerson, C.~Studer, L.~S. Davis,
  G.~Taylor, and T.~Goldstein, ``Adversarial training for free!'' in
  \emph{Proceedings of the 33rd International Conference on Neural Information
  Processing Systems}, 2019, pp. 3358--3369.

\bibitem{Shan2020GottaCA}
S.~Shan, E.~Wenger, B.~Wang, B.~Li, H.~Zheng, and B.~Zhao, ``Gotta catch'em
  all: Using honeypots to catch adversarial attacks on neural networks,''
  \emph{ACM SIGSAC Conference on Computer and Communications Security}, 2020.

\bibitem{sperl2020dla}
P.~Sperl, C.-Y. Kao, P.~Chen, X.~Lei, and K.~B{\"o}ttinger, ``{DLA}:
  dense-layer-analysis for adversarial example detection,'' in \emph{IEEE
  European Symposium on Security and Privacy (EuroS\&P)}, 2020.

\bibitem{Stallkamp2012}
J.~Stallkamp, M.~Schlipsing, J.~Salmen, and C.~Igel, ``Man vs. computer:
  Benchmarking machine learning algorithms for traffic sign recognition,''
  \emph{Neural Networks}, 2012.

\bibitem{steinhardt2017certified}
J.~Steinhardt, P.~W. Koh, and P.~Liang, ``Certified defenses for data poisoning
  attacks,'' in \emph{Proceedings of the 31st International Conference on
  Neural Information Processing Systems}, 2017, pp. 3520--3532.

\bibitem{szegedy2013intriguing}
C.~Szegedy, W.~Zaremba, I.~Sutskever, J.~Bruna, D.~Erhan, I.~J. Goodfellow, and
  R.~Fergus, ``Intriguing properties of neural networks,'' in
  \emph{International Conference on Learning Representations}, 2014.

\bibitem{tian2021detecting}
J.~Tian, J.~Zhou, Y.~Li, and J.~Duan, ``Detecting adversarial examples from
  sensitivity inconsistency of spatial-transform domain,'' in \emph{Proceedings
  of the AAAI Conference on Artificial Intelligence}, vol.~35, no.~11, 2021,
  pp. 9877--9885.

\bibitem{2020}
V.~Venceslai, A.~Marchisio, I.~Alouani, M.~Martina, and M.~Shafique,
  ``Neuroattack: Undermining spiking neural networks security through
  externally triggered bit-flips,'' \emph{International Joint Conference on
  Neural Networks (IJCNN)}, 2020.

\bibitem{wang2019neural}
B.~Wang, Y.~Yao, S.~Shan, H.~Li, B.~Viswanath, H.~Zheng, and B.~Y. Zhao,
  ``Neural cleanse: Identifying and mitigating backdoor attacks in neural
  networks,'' in \emph{IEEE Symposium on Security and Privacy (SP)}, 2019, pp.
  707--723.

\bibitem{wang2014learning}
J.~Wang, Y.~Song, T.~Leung, C.~Rosenberg, J.~Wang, J.~Philbin, B.~Chen, and
  Y.~Wu, ``Learning fine-grained image similarity with deep ranking,'' in
  \emph{Proceedings of the IEEE conference on computer vision and pattern
  recognition}, 2014, pp. 1386--1393.

\bibitem{Wang2020Improving}
Y.~Wang, D.~Zou, J.~Yi, J.~Bailey, X.~Ma, and Q.~Gu, ``Improving adversarial
  robustness requires revisiting misclassified examples,'' in
  \emph{International Conference on Learning Representations}, 2020.

\bibitem{pmlr-v37-xiao15}
H.~Xiao, B.~Biggio, G.~Brown, G.~Fumera, C.~Eckert, and F.~Roli, ``Is feature
  selection secure against training data poisoning?'' in \emph{Proceedings of
  the 32nd International Conference on Machine Learning}, 2015.

\bibitem{xiao2019seeing}
Q.~Xiao, Y.~Chen, C.~Shen, Y.~Chen, and K.~Li, ``Seeing is not believing:
  Camouflage attacks on image scaling algorithms,'' in \emph{28th USENIX
  Security Symposium (USENIX Security 19)}, 2019, pp. 443--460.

\bibitem{XuFeatureSqueezingDetecting2018}
W.~Xu, D.~Evans, and Y.~Qi, ``\BIBforeignlanguage{en}{Feature {{Squeezing}}:
  {{Detecting Adversarial Examples}} in {{Deep Neural Networks}}},'' in
  \emph{\BIBforeignlanguage{en}{Proceedings 2018 {{Network}} and {{Distributed
  System Security Symposium}}}}.\hskip 1em plus 0.5em minus 0.4em\relax {San
  Diego, CA}: {Internet Society}, 2018.

\bibitem{DeepHammer}
F.~Yao, A.~S. Rakin, and D.~Fan, \emph{DeepHammer: Depleting the Intelligence
  of Deep Neural Networks through Targeted Chain of Bit Flips}, 2020.

\bibitem{yaz2018unusual}
Y.~Yaz, C.-S. Foo, S.~Winkler, K.-H. Yap, G.~Piliouras, V.~Chandrasekhar
  \emph{et~al.}, ``The unusual effectiveness of averaging in gan training,'' in
  \emph{International Conference on Learning Representations}, 2018.

\bibitem{zagoruyko2016wide}
S.~Zagoruyko and N.~Komodakis, ``Wide residual networks,'' \emph{arXiv preprint
  arXiv:1605.07146}, 2016.

\bibitem{zhang2019theoretically}
H.~Zhang, Y.~Yu, J.~Jiao, E.~Xing, L.~El~Ghaoui, and M.~Jordan, ``Theoretically
  principled trade-off between robustness and accuracy,'' in
  \emph{International Conference on Machine Learning}.\hskip 1em plus 0.5em
  minus 0.4em\relax PMLR, 2019, pp. 7472--7482.

\bibitem{zhang2021geometryaware}
J.~Zhang, J.~Zhu, G.~Niu, B.~Han, M.~Sugiyama, and M.~Kankanhalli,
  ``Geometry-aware instance-reweighted adversarial training,'' in
  \emph{International Conference on Learning Representations}, 2021.

\bibitem{zhao2020differentiable}
S.~Zhao, Z.~Liu, J.~Lin, J.-Y. Zhu, and S.~Han, ``Differentiable augmentation
  for data-efficient gan training,'' \emph{Advances in Neural Information
  Processing Systems}, vol.~33, 2020.

\end{thebibliography}
